\newif\iffigs\figstrue
\documentclass[12pt]{article}
\usepackage{latexsym,amssymb}
\usepackage[vsentermath]{youngtab}
\iffigs
             \input{epsf}
\else
\fi
\def\IC{\relax\,\hbox{$\inbar\kern-.3em{\rm C}$}}
\def\IG{\relax\,\hbox{$\inbar\kern-.3em{\rm G}$}}
\def\IB{\relax{\rm I\kern-.18em B}}
\def\ID{\relax{\rm I\kern-.18em D}}
\def\IL{\relax{\rm I\kern-.18em L}}
\def\IF{\relax{\rm I\kern-.18em F}}
\def\IH{\relax{\rm I\kern-.18em H}}
\def\II{\relax{\rm I\kern-.17em I}}
\def\IN{\relax{\rm I\kern-.18em N}}
\def\IP{\relax{\rm I\kern-.18em P}}
\def\IQ{\relax\,\hbox{$\inbar\kern-.3em{\rm Q}$}}
\def\bfzero{\relax\,\hbox{$\inbar\kern-.3em{\rm 0}$}}
\def\IK{\relax{\rm I\kern-.18em K}}
\def\IG{\relax\,\hbox{$\inbar\kern-.3em{\rm G}$}}
 \font\cmss=cmss10 \font\cmsss=cmss10 at 7pt
\def\IR{\relax{\rm I\kern-.18em R}}
\def\ZZ{\relax\ifmmode\mathchoice
{\hbox{\cmss Z\kern-.4em Z}}{\hbox{\cmss Z\kern-.4em Z}}
{\lower.9pt\hbox{\cmsss Z\kern-.4em Z}} {\lower1.2pt\hbox{\cmsss
Z\kern-.4em Z}}\else{\cmss Z\kern-.4em Z}\fi}
\def\bfone{\relax{\rm 1\kern-.35em 1}}

\def\inbar{\vrule height1.5ex width.4pt depth0pt}
\def\bfzero{\relax{\rm I\kern-.18em 0}}
\def\bfone{\relax{\rm 1\kern-.35em 1}}

\def\e{\epsilon}

\newcommand{\ft}[2]{{\textstyle\frac{#1}{#2}}}

\def\1bar{1\hskip -.275cm -}
\def\2bar{2\hskip -.275cm -}
\def\3bar{3\hskip -.275cm -}

\newsavebox{\uuunit}
\sbox{\uuunit}
                 {\setlength{\unitlength}{0.825em}
                      \begin{picture}(0.6,0.7)
                                      \thinlines
                                      \put(0,0){\line(1,0){0.5}}
                                      \put(0.15,0){\line(0,1){0.7}}
                                      \put(0.35,0){\line(0,1){0.8}}
                                     \multiput(0.3,0.8)(-0.04,-0.02){10}{\rule{0.5pt}{0.5pt}}
                      \end {picture}}

\newtheorem{teorema}{No go Theorem}[section]

\newtheorem{lemma}{Levi Theorem}[section]

\newtheorem{corollario}{Theorem}[section]
\newtheorem{proofteo}{Proof}[teorema]

\makeatletter \@addtoreset{equation}{section} \makeatother

\newcommand{\be}{\begin{equation}}
\newcommand{\ee}{\end{equation}}
\newcommand{\ba}{\begin{eqnarray}}
\newcommand{\ea}{\end{eqnarray}}
\def\bfone{\relax{\rm 1\kern-.35em 1}}

\def\bfone{\relax{\rm 1\kern-.35em 1}}
\font\cmss=cmss10 \font\cmsss=cmss10 at 7pt

\begin{document}
\begin{titlepage}
\begin{flushright}
DFTT/05/2006\\
{hep-th/0603011}
\end{flushright}
\vskip 0.5cm
\begin{center}
{\LARGE {\bf Twisted tori and  fluxes:
}}\\[0.3cm]
{\LARGE {\bf
a no go theorem for Lie groups of weak $\mathrm{G_2}$ holonomy$^\dagger$}}\\[1.cm]
{\large P. Fr\'e$^{a}$ and M. Trigiante$^{b}$}
{}~\\
\quad \\
{{\em $^{a}$ Dipartimento di Fisica Teorica, Universit\'a di Torino,}}
\\
{{\em $\&$ INFN - Sezione di Torino}}\\
{\em via P. Giuria 1, I-10125 Torino, Italy}~\quad\\
{\tt fre@to.infn.it}
{}~
{}~
{}~\\
\quad \\
{{\em $^{b}$ Dipartimento di Fisica,}}\\
{{\em  Politecnico di Torino, C.so Duca degli Abruzzi,
24,
I-10129 Torino, Italy}}~\quad\\
{\tt trigiante@polito.it}
{}~
{}~
\quad \\
\end{center}
~{}
\begin{abstract}
In this paper we prove the theorem that there exists no
$7$--dimensional Lie group manifold $\mathcal{G}$  of weak
$\mathrm{G_2}$ holonomy.
 We actually prove a stronger statement, namely that there exists
 no $7$--dimensional Lie group with negative definite Ricci tensor ${\bf Ric}_{IJ}$.
 This result rules out (supersymmetric and non--supersymmetric)
 Freund--Rubin solutions of $M$--theory  of the form $\mathrm{AdS_4}\times
 \mathcal{G}$ and compactifications with non--trivial 4--form fluxes of Englert type on an internal group manifold
 $\mathcal{G}$.
A particular class of such backgrounds which, by our arguments are excluded as bulk supergravity compactifications
corresponds to the so called compactifications on twisted--tori, for which $\mathcal{G}$ has structure constants
 $\tau^K{}_{IJ}$ with vanishing trace $\tau^J{}_{IJ}=0$.
On the other hand our result does not have bearing on warped
compactifications of $M$--theory to four dimensions and/or to
compactifications in the presence of localized sources (D--branes,
orientifold planes and so forth). Henceforth our result singles
out the latter compactifications as the preferred hunting grounds
that need to be more systematically explored in relation with all
compactification features involving twisted tori.
\end{abstract}
\vfill
\vspace{1.5cm}
\vspace{2mm} \vfill \hrule width 3.cm {\footnotesize $^ \dagger $
This work is supported in part by the European Union RTN contract
MRTN-CT-2004-005104 and by the Italian Ministry of University (MIUR) under contract PRIN:
named \textit{Simmetrie dell'Universo e delle Interazioni Fondamentali
}}
\end{titlepage}
\section{Introduction}

Recently considerable attention has been devoted to flux
compactifications of super--string theory or of $M$--theory, since
they provide mechanisms to stabilize the moduli fields
\cite{ss}--\cite{df2}. Within this context a particularly
interesting class of flux compactifications is represented by
those on twisted--tori
\cite{ss,km,adfl1,adfl2,dh,kstt,df,alt,ddf,dft1,dft2,dp,dft3}.
This is the contemporary understanding of the Scherk-Schwarz
mechanism \cite{ss} of mass generation from extra dimensions. As
it was explained by Hull and Reid-Edwards \cite{h}, twisted tori
are just Lie group manifolds $\mathcal{G}$ modded by the action of
some discrete subgroup $\Delta \subset \mathcal{G}$ which makes
them compact.
\par
A general pattern only recently elucidated is the relation between
fluxes and gauge algebras. Upon dimensional reduction in the
presence of fluxes  one ends up with some lower dimensional gauged
supergravity and a particularly relevant question is that about
the structure of the gauge Lie algebra in relation with the choice
of fluxes. Some years ago the authors of \cite{gaugedsugrapot1}
introduced the concept of embedding matrix and by means of it
classified all the $\mathcal{N}=8$ supergravity gaugings,
originally studied in \cite{dwn,hull}, were the electric group is
taken to be $\mathrm{SL(8,\mathbb{R})} \subset \mathrm{E_{7(7)}}$.
The very idea of \textit{embedding matrix} proved to be very
fertile and pivotal, yet the classification of
\cite{gaugedsugrapot1} was shown to be incomplete in
\cite{adfl1,adfl2} by proving that the hypothesis that the
electric group be $\mathrm{SL(8,\mathbb{R})}$ could be relaxed. In
the same papers some new explicit examples of gaugings were also
explicitly constructed. Later the authors of \cite{fluxgauge2}
obtained an elegant characterization of the embedding matrix as a
suitable irreducible representation of $\mathrm{E_{7(7)}}$ thus
extending the action of the global symmetry group of the ungauged
theory (also referred to as  $\mathrm{U}$--duality group) to
gauged supergravities. This  analysis was eventually applied to
gauged five--dimensional  maximal supergravities in
\cite{gaugesugrapot2,dwstD5}. Still later, from inspection of the
low--energy description of various flux compactifications (either
involving form--fluxes or ``geometrical fluxes'', i.e. background
quantities related to the geometry of the internal manifold, like
the ``twist'' tensor characterizing the twisted--tori), a precise
statement about the relation between internal fluxes and local
symmetries of the lower--dimensional effective supergravity was
derived: the background quantities enter the lower--dimensional
gauged supergravity as components of the embedding tensor, and
thus they can be naturally assigned to representations of the
$\mathrm{U}$--duality group \cite{alt} (a similar statement was
made in the context of the heterotic string in \cite{km}).
 Hence an obvious question is that relative to the gauge algebra emerging from
flux compactifications on twisted tori. This question has been
addressed in a recent series of papers \cite{df,ddf,dft1,dft2} and
it has been advocated that the gauge algebraic structures emerging
in $D=4$ supergravities which originate  from this kind of
$M$--theory compactifications, do not fall in the class of Lie
algebras $\mathbb{G}$, but rather have to be understood in the
more general context of Free Differential Algebras. The algebraic
structure that goes under this name was independently discovered
at the beginning of the eighties in Mathematics by Sullivan
\cite{sullivan} and in Physics by the authors of \cite{fredauria}.
Free Differential Algebras (FDA) are a categorical extension of
the notion of Lie algebra and constitute the natural mathematical
environment for the description of the algebraic structure of
higher dimensional supergravity theory, hence also of string
theory. The reason is the ubiquitous presence in the spectrum of
string/supergravity theory of antisymmetric gauge fields
($p$--forms) of rank greater than one. The very existence of FDA.s
is a consequence of Chevalley cohomology of ordinary Lie algebras
and Sullivan has provided us with a very elegant classification
scheme of these algebras based on two structural theorems rooted
in the set up of such an elliptic complex.
\par
In view of Sullivan's theorems one of the present authors analyzed in
\cite{chevrolet} the results of
 \cite{df,ddf,dft1,dft2}
from the point of view of Chevalley cohomology. The goal was that of
establishing the structure of the minimal FDA algebra $\mathbb{M}$
which emerges from twisted tori compactifications and relating its
\textit{generalized structure constants} to the fluxes, just  in the spirit
of the general relation between \textit{gaugings}, i.e. embedding
matrices and $p$-form fluxes explained above.  The basic notion of
\textit{minimal algebra} is illustrated in the following way.
\par
As it was noted in \cite{comments},  FDA.s have the additional
fascinating property that, as opposite to ordinary Lie algebras,
they already encompass their own gauging. Indeed the first of
Sullivan's structural theorems, which is in some sense analogous
to Levi's theorem for Lie algebras, states that the most general
FDA is a semidirect sum of a so called minimal algebra
$\mathbb{M}$ with a contractible one $\mathbb{C}$. The generators
of the minimal algebra are physically interpreted as the
connections or \textit{potentials}, while the contractible
generators are physically interpreted as the \textit{curvatures}.
The real hard--core of the FDA is the minimal algebra and it is
obtained by setting the contractible generators (the curvatures)
to zero. The structure of the minimal algebra $\mathbb{M}$, in
turn, is beautifully determined by Chevalley cohomology of the
underlying Lie algebra  $\mathbb{G}$. This happens to be the
content of Sullivan's second structural theorem.
\par
By rephrasing all the equations of papers \cite{df,ddf,dft1,dft2}
in the framework of Chevalley cohomology and exploiting the
underlying structure of a double elliptic complex,  paper
\cite{chevrolet} established the following result:
\begin{corollario}\label{teor1}
The minimal algebra $\mathbb{M}$ emerging in twisted tori
compactification of M-theory, i.e. in compactifications on a $7$-dimensional group manifold
 $\mathcal{G}$
 necessarily coincides with the  Lie algebra
$\mathbb{G}_7$ of $\mathcal{G}$ unless the internal flux
$g_{IJKL}$ of the $4$-form defines a cohomologically non-trivial
$4$-cycle of $\mathbb{G}_7$, namely unless  $g_{IJKL} e^I \wedge
e^J \wedge e^K \wedge e^L \equiv \Delta^{(0,4)} \in
\mathrm{H^{(4)}}(\mathbb{G}_7)$.
\end{corollario}

It was actually proven in \cite{chevrolet} that  in order to get a
 minimal algebra which is a proper extension of $\mathbb{G}_7$ on a certain background, $\Delta^{(0,4)} \in
\mathrm{H^{(4)}}(\mathbb{G}_7)$ is just a necessary but not yet
sufficient condition. Indeed it was proven the following
\begin{corollario}\label{teor2}
The necessary and sufficient condition for the minimal free
differential algebra $\mathbb{M}$ to be a proper extension of
$\mathbb{G}_7$, is that the $4$-form $\Delta^{(0,4)}$ defined by
the internal $4$--form flux should:
\begin{description}
  \item[a] be cohomologically non trivial $\Delta^{(0,4)} \in
  \mathrm{H^{(4)}}(\mathbb{\mathbb{G}}_7)$
  \item[b] its triple contraction should be a non trivial $1$--cycle.
  Given a basis of cycles $\Gamma^{[p]}$ for each Chevalley cohomology
  group $H^{(p)} (\mathbb{G}_7)$, and the pairing form $< \, , \, >$, this condition is expressed by:
  \begin{equation}
  \exists \,\Gamma_\alpha^{[6]} \, \in \, \mathrm{H^{(6)}} (\mathbb{G}_7) \quad \backslash
  \quad 0  \, \ne \, \langle i_W\circ i_W \circ
  i_W \, \Delta^{[0,4]} \, , \, \Gamma_\alpha^{[6]} \rangle
\label{bellet}
\end{equation}
\end{description}
\end{corollario}
Hence paper \cite{chevrolet} reduced the problem of establishing
whether or not the minimal part of the FDA.s,  that emerge from
twisted tori compactifications of $M$--theory, involve forms other
than the Kaluza--Klein vectors $W^I_\mu$,  to the following three
questions about $7$-dimensional Lie algebras $\mathbb{G}_7$:
\begin{description}
  \item [A] Do $7$-dimensional Lie algebras $\mathbb{G}_7$ exist which admit  a
  constant rank $4$ antisymmetric tensor $g_{IJKL}\, \in \, \bigwedge^{(4)} \, \mathrm{adj} \mathbb{G}_7$
  which together with the Ricci $2$-form of the same algebra
  $\mathbf{Ric}(\mathbb{G}_7)$ provides an exact solution of
  $M$--theory field equations?
  \item [B] If the answer to   question [A] is yes, can $\Delta^{[0,4]} \equiv g_{IJKL}
  e^I \wedge e^J \wedge e^K \wedge e^L $ be chosen cohomologically
  non trivial, namely within $\mathrm{H^{(4)}} (\mathbb{G}_7)$ ?
  \item [C] If the answer to questions [A] and [B] is yes, can the
  $4$-cycle $\Delta^{[0,4]}$ be chosen in such a way that it also
  satisfies the condition (\ref{bellet})?
\end{description}
In the present paper by using general classical theorems in Lie
algebra theory we shall prove that  the answer to question [A] is
already negative so that questions [B] and [C] need not even be
addressed. This is the no-go theorem announced by our title. Let
us however immediately comment on the scope and bearings of our
theorem. Indeed, we are quite conscious that the value of any
negative result resides in a clear--cut presentation of the
hypotheses and conditions under which it is derived. Only in this
case the negative statement can be turned into a positive
contribution by indicating the directions to be pursued in order
to evade it.
\par
So let us immediately recall that by compactification of  M-theory
on a twisted torus we mean in this paper the same that was meant
in \cite{chevrolet} and that was assumed in the series of papers
\cite{df,ddf,dft1,dft2}, namely a bosonic configuration of the
$M$--theory light fields (the metric $g_{{\hat \mu}{\hat \nu}}$
and the three form $A_{{\hat \mu}{\hat \nu}{\hat \rho}}$) such
that the $11$--dimensional manifold splits into the following
direct product:
\begin{equation}
  \mathcal{M}_{11} = \mathcal{M}_4 \, \times \, \mathcal{G}/\Delta
\label{M11split}
\end{equation}
$\mathcal{M}_4$ denoting a four-dimensional maximally symmetric
manifold whose coordinates we denote $x^\mu$ and $\mathcal{G}$ a
$7$--dimensional group manifold whose parameters we denote $y^I$.
Let us denote by
\begin{equation}
  e^I= e^I_J(y) \, dy^J
\label{linvforme}
\end{equation}
the purely $y$-dependent left--invariant $1$-forms on $\mathcal{G}$
which, by definition, satisfy the Maurer--Cartan equations:
\begin{equation}
  \partial e^I = \ft 12 \, \tau^{I}_{\phantom{I}JK} \, e^J \wedge
  e^K\,\,;\,\,\,I,J,K=4,\dots\ 10
\label{MC1}
\end{equation}
$\tau^{I}{}_{JK}$ being the structure constants of the Lie algebra
$\mathbb{G}_7$:
\begin{equation}
  \left[ T_I \, , T_J\right] \, = \, \tau^{K}_{\phantom{I}IJ} \,
  T_K
\label{commrel}
\end{equation}
It is a constitutive part of what we understand by
compactification that in any configuration of the compactified
theory the eleven dimensional vielbein is  split as follows:
\begin{equation}
  V^{\hat a} =\left \{\begin{array}{rclcrcl}
    V^r & = & E^r (x) & ; & r & = & 0,1,2,3 \\
    V^a & = & \Phi^a_{\phantom{I}J}(x) \, \left( e^I+W^I(x)\right)  & ; & a & = & 4,5,6,7,8,9,10 \
  \end{array} \right.
\label{conv1}
\end{equation}
where $E^r(x)$ is a purely $x$--dependent $4$--dimensional
vielbein, $W^I(x)$ is an $x$--dependent $1$--form on $x$-space
describing the Kaluza Klein vectors and the purely $x$--dependent
$7 \times 7$ matrix $\Phi^a_{\phantom{I}J}(x)$ encodes part of the
scalar fields of the compactified theory, namely the internal
metric moduli. At the same time the $3$-form is expanded as
follows:
\begin{eqnarray}
\mathbf{A}^{[3]} & = & C^{[0]}_{IJK}(x) \, V^I \wedge V^J \wedge V^K + A^{[1]}_{IJ}(x) \wedge V^I \wedge V^J
  \, + \, B^{[2]}_I(x) \, \wedge \, V^I + {A}^{[3]}(x)\nonumber\\
\label{3f1}
\end{eqnarray}
where $V^I=e^I+W^I(x)$, $ C^{[0]}_{IJK}(x)$ are $x$-dependent
scalar fields, $A^{[1]}_{IJ}(x)$ are $x$-dependent $1$-forms
$B^{[2]}_I(x)$ are $x$-dependent $2$-forms and finally
${A}^{[3]}(x)$ is an $x$--dependent $3$-form. All of these forms
are assumed to live on $D=4$ space-time. From these assumptions it
follows that the bosonic field strength is expanded as follows:
\begin{eqnarray}
{ {\mathbf{F}}}^{[4]} & \equiv & F^{[4]}(x) \,
 + \, F^{[3]}_I(x) \,\wedge \,  V^I + \, F^{[2]}_{IJ}(x) \,\wedge \,  V^I \,\wedge
\,V^J \nonumber\\
 && + \, F^{[1]}_{IJK}(x) \,\wedge \,  V^I \,\wedge \,V^J \, \wedge \,
 V^K \, + \,  F^{[0]}_{IJKL}(x) \,\wedge \,  V^I \,\wedge \,V^J \, \wedge \,
 V^K \, \wedge \, V^L
\label{fb1}
\end{eqnarray}
where $F^{[p]}_{I_{1}\dots I_{4-p}}(x)$ are $x$-space $p$--forms
depending only on $x$.
\par
In bosonic backgrounds with space--time geometry (\ref{M11split}),
the
family of configurations (\ref{conv1}) must satisfy the condition
that by choosing:
\begin{eqnarray}
E^r & = & \mbox{vielbein of a maximally symmetric $4$-dimensional
space time
} \label{maxsym}\\
\Phi^I_{\phantom{I}J}(x) & = & \delta^{I}_{\phantom{I}J} \label{delta=Phi}\\
W^I & = & 0 \label{noKK}\\
F^{[3]}_I(x) & = & F^{[2]}_{IJ}(x) \, = \, F^{[1]}_{IJK}(x) \, = \,
0 \label{viaperlorentz}\\
 F^{[4]}(x) & = & e \, \epsilon_{rstu} \, E^r \, \wedge \, E^s \,
 \wedge \, E^t \, \wedge \, E^u  \quad ; \quad (e = \mbox{constant parameter})\label{freundrub}\\
F^{[0]}_{IJKL}(x) & = & g_{IJKL} \, = \, \mbox{constant tensor}
\label{intoflux}
\end{eqnarray}
we obtain an exact \textit{bona fide} solution of the
eleven--dimensional field equations of M-theory.
\par
There are three possible eleven--dimensional backgrounds of this
kind:
\begin{equation}
  \mathcal{M}_4 \, = \, \left\{ \begin{array}{cc}
    \mathcal{M}_4 & \mbox{Minkowsky space} \\
    \mathrm{dS_4} & \mbox{de Sitter space} \\
    \mathrm{AdS_4} & \mbox{anti de Sitter space} \
  \end{array} \right.
\label{threepossib}
\end{equation}
In any case Lorentz invariance imposes
eqs.(\ref{delta=Phi},\ref{noKK},\ref{viaperlorentz}) while
translation invariance imposes that the vacuum expectation value
of the scalar fields $\Phi^I_{\phantom{I}J}(x)$ should be a
constant matrix
\begin{equation}
 < \Phi^I_{\phantom{I}J}(x) > = \mathcal{A}^{I}_{J}
\label{Amatra}
\end{equation}
while the tensor $g_{IJKL}$  should be a constant tensor as assumed
in eq.(\ref{intoflux}). The reason why we set
$\mathcal{A}^{I}_{J}=\delta^{I}_{\phantom{I}J}$ is that any matrix
$\mathcal{A}$ can be reabsorbed by a change of basis of generators of
the Lie algebra and hence it corresponds to no new degrees of
freedom.
\par
In this paper we shall prove that there are no $\mathbb{G}_7$ Lie
algebras such that eq.s (\ref{maxsym}-\ref{intoflux}) lead to a
bona fide solution of M-theory field equations with a
cohomologically non--trivial flux $g_{IJKL}$. This, in view of our
previous discussion, implies the no go theorem on FDA claimed
above. The same result was obtained in \cite{dft2,dft4} from
inspection of the scalar potential $\mathcal{V}$ of the
dimensionally reduced theory. Indeed bosonic backgrounds of the
kind we are considering, correspond to extrema of $\mathcal{V}$.
It was shown that the extremization of $\mathcal{V}$ with resect
to the axions $C_{IJK}$ implies the vanishing of the internal
components $F^{[0]}_{IJKL}$ of the 4--form field strength:
\begin{eqnarray}
F^{[0]}_{IJKL}&=& -g_{IJKL}-\frac{3}{2}\tau^M{}_{[IJ}C_{KL]M}=0
\end{eqnarray}
which in turn implies that $g_{IJKL}$ is cohomologically trivial.
Still from inspection of $\mathcal{V}$ in \cite{dft4}, solutions
with de Sitter geometry ($\mathcal{V}>0$ at the minimum) were
ruled out, but no statement was made about the existence of
anti--de Sitter vacua. In \cite{dp} the existence of
supersymmetric anti--de Sitter vacua was ruled out in a particular
orientifold truncation of the maximally supersymmetric model. The
output of our present analysis is two--fold: on  one hand we give
an alternative  formal derivation of the triviality of $g_{IJKL}$,
making use of the aforementioned arguments, on the other hand we
show that bosonic backgrounds of anti--de Sitter type (arising
from Freund --Rubin type of compactifications) are not solutions
of the theory.
\par
It should be noted that there are two main assumptions which could be
possibly relaxed  leading, may be, to different conclusions:
\begin{enumerate}
  \item We could introduce a warp factor. This means that in eq.
  (\ref{conv1}) the first line could be replaced by
\begin{equation}
  V^a = \exp[\mathcal{W}(y)] \, E^a(x)
\label{warping}
\end{equation}
where $\mathcal{W}(y)$ is some suitable function of the internal
coordinates. In this case, however, the entire analysis of eq.s
(\ref{3f1}) and (\ref{fb1}) in terms of Chevalley cohomology has to
be reconsidered since the internal boundary operator $\partial$,
defining such a cohomology, no longer annihilates the $D=4$
vielbeins $(\partial V^a \ne 0)$.
  \item Our result applies to the field equations of M-theory without
  sources. Including the contribution of $M2$ or $M5$ branes one
  changes those equations by the addition of new terms whose bearing has to
  be considered.
\end{enumerate}
The token by means of which we shall prove our negative result is
provided by the reduction of the whole problem to a question of
holonomy on the considered internal manifold $\mathcal{M}_7$ (in
our case the group manifold $\mathcal{G}$). Indeed as we explain
in detail in section \ref{englertspace} a solution of M-theory
field equations with a non--trivial internal flux does exist if
and only if the internal $7$-manifold $\mathcal{M}_7$ is a
Riemanian manifold of \textit{weak $\mathrm{G_2}$
holonomy}\cite{weakG2}. This is the new phrasing, in modern
parlance, of the notion originally introduced at the beginning of
the eighties by the authors of \cite{fermionMassSpectrum} under
the name of \textit{Englert manifolds}.
\par
In view of this the final question will be whether $7$-dimensional
group manifolds of \textit{weak $\mathrm{G_2}$ holonomy} do or do
not exist. Compact coset manifolds $\mathcal{S}/\mathcal{R}$  of
the Englert type (weak $\mathrm{G_2}$ holonomy) were all
classified and studied in the golden age of Kaluza Klein
supergravity namely at the beginning of the eighties\footnote{See
\cite{castdauriafre} for a comprehensive review and the basic
original literature for the various families of
$\mathcal{S}/\mathcal{R}$ spaces namely
\cite{freundrubin,roundseven,roundspectrum} for $\mathbb{S}^7$,
\cite{Awada:1982pk} for the squashed $\mathbb{S}^7$,
\cite{su3su2u1} for the $M^{pqr}$ spaces, \cite{D'Auria:1983vy}
for $Q^{pqr}$ spaces, \cite{Castellani:1983tc} for the $N^{pqr}$
spaces and \cite{classificoleo} for the exhaustive classification
of the remaining ones.}. Yet group manifolds, possibly
non--compact, have not been considered  so far under this point of
view. We shall be able to prove the:
\begin{corollario}\label{nogruppo}
 $7$-dimensional group manifolds of Englert type, namely of weak
$\mathrm{G_2}$ holonomy do not exist.
\end{corollario}
Using the interesting results of \cite{weakG2} we will reformulate
the condition of weak $\mathrm{G_2}$ holonomy in terms of the
irreducible $\mathrm{G_2}$ representation contents of the
candidate structure constants $\tau^{I}_{\phantom{I}JK} $ of the
Lie algebra $\mathbb{G}_7$. This leads to a parametrization of
$\tau^{I}_{\phantom{I}JK}$ in terms of $91$--parameters. We should
now impose Jacobi indentities and find how many solutions survive.
This turns out to be algebraically too difficult. Yet we can use a
different powerful argument. If a Lie algebra of weak
$\mathrm{G_2}$ holonomy existed its Ricci tensor ${\bf
Ric}^I{}_J$\footnote{In our conventions the curvature tensor is
 $-1/2$ times the definition which is traditionally adopted in
the literature on General Relativity, so that for instance the
Ricci tensor ${\bf Ric}_{IJ}$ of a sphere is negative definite.
Moreover we use the ``mostly minus'' convention for the metric, so
that the internal manifold has a negative definite metric.} would
be proportional to the identity matrix with a \emph{positive}
coefficient. We can instead prove that for all possible
$7$-dimensional Lie algebras the Ricci form ${\bf Ric}^I{}_J$ has
an indefinite signature admitting at least one negative or null
eigenvalue.
\par
Our paper is organized as follows
\par
In section \ref{englertspace} we analyze $M$-theory field
equations and the notion of weak $\mathrm{G_2}$ holonomy. We show
that this is identical to the notion of \textit{Englert manifold}
introduced in the literature on Kaluza Klein supergravity two
decades ago and we demonstrate that this is the necessary and
sufficient condition in order to find solutions with an internal
flux.
\par
In section \ref{grupporiformulo} we reformulate the notion of weak
$\mathrm{G_2}$ holonomy in terms of representation theory.
\par
In section \ref{ricciussection} we prove our main no-go theorem.
\par
Section \ref{concludo} briefly summarizes our conclusions.
\par
Appendix \ref{g2formalismo} contains several details on $\mathrm{G_2}$
invariant forms, projection operators and irreducible tensors.
\par
Appendix \ref{Bappendo} contains a classification of seven
dimensional algebras with either $\mathrm{SO(3)}$ or $\mathrm{SO(1,2)}$ Levi
subalgebras. These classifications are utilized in the proof of our main no-go
theorem.
\section{M-theory field equations and $7$-manifolds of weak $\mathrm{G_2}$
holonomy \textit{i.e.} Englert $7$-manifolds}
\label{englertspace}
\par
In the recent literature about flux compactifications a geometrical notion which
has been extensively exploited is that of $\mathrm{G}$-structures.
\par
Following, for instance, the presentation of \cite{tommasiello},
if $\mathcal{M}_n$ is a differentiable manifold of dimension $n$,
$T\mathcal{M}_n \stackrel{\pi}{\rightarrow} \mathcal{M}_n $ its
tangent bundle and $F\mathcal{M}_n \stackrel{\pi}{\rightarrow}
\mathcal{M}_n $ its frame bundle, we say that $\mathcal{M}_n$
admits a $\mathrm{G}$-structure when the structural group of
$F\mathcal{M}_n$ is reduced from the generic
$\mathrm{GL(n,\mathbb{R})}$ to a proper subgroup $\mathrm{G}
\subset \mathrm{GL(n,\mathbb{R})}$. Generically, tensors on
$\mathcal{M}_n$ transform in representations of the structural
group $\mathrm{GL(n,\mathbb{R})}$. If a $\mathrm{G}$-structure
reduces this latter to $\mathrm{G} \subset
\mathrm{GL(n,\mathbb{R})}$, then the decomposition of an
 irreducible representation of $\mathrm{GL(n,\mathbb{R})}$, pertaining
to a certain tensor $t^{p}$, with respect to the subgroup
$\mathrm{G}$ may contain singlets. This means that on such a
manifold $\mathcal{M}_n$ there may exist a certain tensor $t^{p}$
which is $\mathrm{G}$--invariant, and therefore globally defined.
As recalled in \cite{tommasiello} existence of a Riemannian metric
$g$ on $\mathcal{M}_n$ is equivalent to a reduction of the
structural group $\mathrm{GL(n,\mathbb{R})}$ to $\mathrm{O(n)}$,
namely to an $\mathrm{O(n)}$-structure. Indeed, one can reduce the
frame bundle by introducing orthonormal frames, the vielbein
$e^I$, and, written in these frames, the metric is the the
$\mathrm{O(n)}$ invariant tensor $\delta_{IJ}$.  Similarly
orientability corresponds to an $\mathrm{SO(n)}$-structure and the
existence of spinors on spin manifolds corresponds to a
$\mathrm{Spin}(n)$-structure.
\par
In the case of seven dimensions, an orientable Riemannian manifold
$\mathcal{M}_7$, whose frame bundle has generically an
$\mathrm{SO(7)}$ structural group admits a
$\mathrm{G_2}$-structure if and only if, in the basis provided by
the orthonormal frames $e^I$, there exists an  antisymmetric
$3$-tensor $\phi_{IJK}$ satisfying the algebra of the octonionic
structure constants:
\begin{eqnarray}
  \phi_{ABK} \, \phi_{CDK} & = & \ft {1}{18} \, \delta^{CD}_{AB} \,
  - \, \ft {2}{3} \, \phi^{\star}_{ABCD} \nonumber\\
  - \ft {1}{6} \, \epsilon_{IJKABCD} \, \phi^{\star}_{ABCD} & = & \phi_{IJK}
\label{octonio}
\end{eqnarray}
which is invariant, namely it is the same in all local
trivializations of the $\mathrm{SO(7)}$ frame bundle.
This corresponds to the algebraic definition of
$\mathrm{G_2}$ as that subgroup of $\mathrm{SO(7)}$ which acts as
an automorphism group of the octonion algebra.
Alternatively $\mathrm{G_2}$ can be defined as the stability subgroup of the
$8$-dimensional spinor representation of $\mathrm{SO(7)}$.  Hence
we can equivalently state that a manifold $\mathcal{M}_7$ has a
$\mathrm{G_2}$-structure if there exists at least an invariant
spinor $\eta$, which is the same in all local trivializations of the
$\mathrm{Spin(7)}$ spinor bundle.
\par
In terms of
this invariant  spinor the invariant 3--tensor $\phi_{IJK}$ has the form
\footnote{For more details on the $\mathrm{G_2}$-formalism see appendix \ref{g2formalismo}}:
\begin{equation}
  \phi_{IJK} = \ft 16 \, \eta^T \, \gamma_{IJK} \, \eta
\label{etaTeta}
\end{equation}
and eq.(\ref{etaTeta}) provides the relation between the two
definitions of the $\mathrm{G_2}$-structure.
\par
On the other hand the manifold has not only a $\mathrm{G_2}$--structure, but also  $\mathrm{G_2}$--holonomy
if the  invariant three--tensor $\phi_{ABK}$ is covariantly constant. Namely we must have:
\begin{eqnarray}
0 & = & \nabla \phi_{ABK} \, \equiv \, d\phi_{ABK} \, + \, 3 \, \omega_{P[I} \,  \phi_{JK]P}
\label{octonioHOLO}
\end{eqnarray}
where the $1$-form $\omega^{AB}$ is the spin connection of $\mathcal{M}_7$.
Alternatively the manifold has $\mathrm{G_2}$--holonomy if the
invariant spinor $\eta$ is covariantly constant, namely if:
\begin{equation}
  \exists \, \eta \, \in \, \Gamma(\mathrm{Spin} \mathcal{M}_7 ,
  \mathcal{M}_7) \quad \backslash \quad  0 \, = \,\nabla \, \eta \, \equiv d\eta
  -\ft 14 \, \omega^{IJ} \, \gamma_{IJ} \, \eta
\label{covspinor}
\end{equation}
where $\gamma_I$ ($I =1,\dots, 7$) are the $8 \times 8$ gamma matrices of the $\mathrm{SO(7)}$
Clifford algebra.
The relation between the two definitions (\ref{octonioHOLO}) and
(\ref{covspinor}) of  $\mathrm{G_2}$-holonomy  is the same as for the
two definitions of the $\mathrm{G_2}$-structure, namely it is given by eq.(\ref{etaTeta}).
As a consequence of its own definition a Riemannian $7$-manifold with
$\mathrm{G_2}$ holonomy  is Ricci flat. Indeed the integrability condition of eq.(\ref{covspinor})
yields:
\begin{equation}
  \mathcal{R}^{AB}_{JK} \, \gamma_{AB} \, \eta \, = \, 0
\label{holonomia1}
\end{equation}
where $\mathcal{R}^{AB}_{JK}$ is the Riemann tensor of
$\mathcal{M}_7$. From eq.(\ref{holonomia1}), by means of a few simple
algebraic manipulations one obtains two results:
\begin{itemize}
  \item{ The curvature $2$-form
\begin{equation}
  \mathcal{R}^{AB}\, \equiv \, \mathcal{R}^{AB}_{JK} \, e^J \, \wedge
  \, e^K
\label{curvadue}
\end{equation}
is $\mathrm{G_2}$ Lie algebra valued,  namely it satisfies the condition\footnote{See appendix \ref{g2formalismo}
for details on $\mathrm{G_2}$ decomposition
of $\mathrm{SO(7)}$ representations}:
\begin{equation}
  \phi^{IAB} \, \mathcal{R}^{AB} \, = \, 0
\label{G2condition}
\end{equation}
which projects out the $\mathbf{7}$ of $\mathrm{G_2}$ from the $\mathbf{21}$
of $\mathrm{SO(7)}$ and leaves with the adjoint $\mathbf{14}$.}
  \item {The internal Ricci tensor is zero:
\begin{equation}
  \mathcal{R}^{AM}_{BM} \, =\, 0
\label{riccipiatto}
\end{equation}
   }
\end{itemize}
Next we consider the field equations of $M$-theory \cite{cremmerjulia}, which we write
here  with the notations and the conventions of \cite{fredauria11}
\begin{eqnarray}
0   & = & \mathcal{D}_{\hat m} \, F^{\hat{m}\hat{c}_1 \hat{c}_2 \hat{c}_3} + \frac{1}{96} \,
\epsilon^{\hat{c}_1\hat{c}_2\hat{c}_3\hat{a}_1 \dots \hat{a}_8}
 F_{\hat{a}_1 \dots \hat{a}_4}\, F_{\hat{a}_5 \dots \hat{a}_8}\label{fresconiA}\\
{R}^{\hat{a}\hat{m}}_{\phantom{am}\hat{b}\hat{m}} & =  & 6
F^{\hat{a}\hat{c}_1\hat{c}_2\hat{c}_3} \, F_{\hat{b}\hat{c}_1\hat{c}_2\hat{c}_3} \, - \, \ft
12 \delta^{\hat{a}}_{\hat{b}} \, F^{\hat{c}_1\hat{c}_2\hat{c}_3\hat{c}_4} \, F_{\hat{c}_1\hat{c}_2\hat{c}_3\hat{c}_4}
\label{fresconiB}
\end{eqnarray}
where hatted indices run on eleven values and are flat indices. We
make the compactification ansatz:
\begin{equation}
  \mathcal{M}_{11} =  \mathcal{M}_{7}\, \times \, \mathcal{M}_4
\label{compatta1}
\end{equation}
where $\mathcal{M}_4$ is one of the three possibilities mentioned
in eq.(\ref{threepossib}) and all of
eq.s(\ref{delta=Phi}-\ref{intoflux}) hold true.  Then we split the
rigid index range as follows:
\begin{equation}
  \hat{a},\hat{b},\hat{c},\dots = \cases{
  a,b,c,\dots \, \, = 4,5,6,7,8,9,10  \, = \mbox{$\mathcal{M}_7$ indices}\cr
  r,s,t,\dots  \quad = 0,1,2,3 \quad\quad =
  \mbox{$\mathcal{M}_4$ indices} \cr}
\label{indexsplit}
\end{equation}
and by following the conventions employed in \cite{su3su2u1} and using the
results obtained in the same paper, we conclude that  the compactification ansatz
reduces the system (\ref{fresconiA},\ref{fresconiB})
to the following one:
\begin{eqnarray}
{R}^{rs}_{\phantom{ab}tu} & = & \lambda \, \delta^{rs}_{tu} \label{AdScurva}\\
\mathcal{R}^{IK}_{JK} & = & 3 \, \nu \, \delta^{I}_{J} \label{M7curva}\\
F_{rstu} & = & e \, \epsilon_{rstu} \label{externalflux}\\
g_{IJKL} & = &f\, \mathcal{F}_{IJKL}  \label{internalflux}\\
\mathcal{F}^{AIJK} \, \mathcal{F}_{BIJK} & = & \mu \, \delta^{A}_{B} \label{gsquare}\\
\mathcal{D}^M \, \mathcal{F}_{MIJK} & = & \ft 12 \, e \, \epsilon_{IJKPQRS} \,
\mathcal{F}^{PQRS}\label{englert}
\end{eqnarray}

recall that we use capital latin indices to label the internal
coordinates. Eq. (\ref{M7curva}) states that the internal manifold
$\mathcal{M}_{7}$ must be an Einstein space. Eq.s
(\ref{externalflux}) and (\ref{internalflux}) state that there is
a flux of the four--form both on $4$--dimensional space-time
$\mathcal{M}_4$ and on the internal manifold $\mathcal{M}_7$. The
parameter $e$, which fixes the size of the flux on the
four--dimensional space and was already introduced in
eq.(\ref{freundrub}), is called the Freund-Rubin parameter
\cite{freundrubin}. As we are going to show, in the case that a
non vanishing $\mathcal{F}^{AIJK}$ is required to exist, eq.s
(\ref{gsquare}) and (\ref{englert}),  are equivalent to the
assertion that the manifold $\mathcal{M}_7$ has weak
$\mathrm{G_2}$ holonomy rather than $\mathrm{G_2}$--holonomy, to
state it in modern parlance \cite{weakG2}. In paper
\cite{fermionMassSpectrum}, manifolds admitting such a structure
were instead named \textit{Englert spaces} and the underlying
notion of weak $\mathrm{G_2}$ holonomy was already introduced
there with the different name of \textit{ de Sitter
$\mathrm{SO(7)}^+$ holonomy}.
\par

Indeed eq.(\ref{englert}) which, in the language of the early
eighties was named Englert equation \cite{englertsolution} and
which is nothing else but equation (\ref{fresconiA}), upon
substitution of the Freund Rubin ansatz (\ref{externalflux}) for
the external flux, can be recast in the following more revealing
form: Let
\begin{equation}
  \Phi^\star \, \equiv \, \mathcal{F}_{IJKL} \, e^I \, \wedge \, e^J
  \, \wedge \, \e^K  \, \wedge \, e^L
\label{phistarra}
\end{equation}
be a the constant $4$--form on $\mathcal{M}_7$ defined by our non
vanishing flux, and let
\begin{equation}
  \Phi \, \equiv \,\ft {1}{24} \epsilon_{ABCIJKL} \, \mathcal{F}_{IJKL} \, e^A \, \wedge \,
  e^B
  \, \wedge \, \e^C
\label{nophistarra}
\end{equation}
be its dual. Englert eq.(\ref{englert}) is just the same as writing:
\begin{eqnarray}
  d \Phi & = & 12\, e \, \Phi^\star \nonumber\\
   d \Phi^\star & = & 0
\label{ollaolla}
\end{eqnarray}
When the Freund Rubin parameter vanishes $e=0$ we recognize in
eq.(\ref{ollaolla}) the statement that our internal manifold
$\mathcal{M}_7$ has $\mathrm{G_2}$-holonomy and hence  it is Ricci
flat.  Indeed $\Phi$ is the $\mathrm{G_2}$ invariant and
covariantly constant form defining $\mathrm{G_2}$-structure and
$\mathrm{G_2}$-holonomy. On the other hand the case $e\ne 0$
corresponds to the  weak $\mathrm{G_2}$ holonomy. Just as we
reduced the existence  of a closed three-form $\Phi$ to the
existence of a $\mathrm{G_2}$ covariantly constant spinor
satisfying eq.(\ref{covspinor}) which allows to set the
identification (\ref{etaTeta}), in the same way eq.s
(\ref{ollaolla}) can be solved \textit{if and only if} on
$\mathcal{M}_7$ there exist a weak Killing spinor $\eta$
satisfying the following defining condition:
\begin{equation}
  \mathcal{D}_I \, \eta \, = \, m \, e \, \gamma_I \, \eta
\label{weakkispinor}
\end{equation}
 where $m$ is a numerical constant
and $e$ is the Freund-Rubin parameter, namely the only scale which
at the end of the day will occur in the solution.\par The
integrability of the above equation implies that the Ricci tensor
be proportional to the identity, namely that the manifold is an
Einstein manifold and furthermore fixes the proportionality
constant:
\begin{equation}
  \mathcal{R}^{IM}_{\phantom{IM}JM} = 12 \, m^2 \, e^2 \, \delta^{I}_{J} \quad
  \longrightarrow \, \nu = \, 12 \, m^2 \, e^2
\label{fixingnu}
\end{equation}
In case such a spinor exist, by setting:
\begin{equation}
  g_{IJKL} = \mathcal{F}_{IJKL} = \eta^T \, \gamma_{IJKL} \eta \, = \, 24 \,
  \phi^\star_{IJKL}
\label{agnisco1}
\end{equation}
we find that Englert equation (\ref{englert}) is satisfied, provided
we have:
\begin{equation}
  m = - \frac{3}{2}
\label{tremezzi}
\end{equation}
In this way Maxwell equation, namely
(\ref{fresconiA}) is solved.
Let us also note, as the authors of \cite{fermionMassSpectrum} did
many years ago, that condition (\ref{weakkispinor}) can also be
interpreted in the following way. The spin-connection $\omega^{AB}$
plus the vielbein $e^C$ define on any non Ricci flat $7$-manifold  $\mathcal{M}_7$ a
connection which is actually $\mathrm{SO(8)}$ rather than $\mathrm{SO(7)}$ Lie algebra
valued. In other words we have a principal $\mathrm{SO(8)}$ bundle
which leads to an $\mathrm{SO(8)}$ spin bundle of which $\eta$ is a
covariantly constant section:
\begin{equation}
 0 \, = \,  \nabla^{\mathrm{SO(8)}}\eta = \nabla^{\mathrm{SO(7)}} \,-\,  m \, e \, e^I\,  \gamma_I \,
  \eta
\label{so8cov}
\end{equation}
The  existence of $\eta$  implies a reduction of the $\mathrm{SO(8)}$-bundle.
Indeed the stability subgroup of an $\mathrm{SO(8)}$ spinor is a well
known subgroup $\mathrm{SO(7)}^+$ different from the standard $\mathrm{SO(7)}$ which,
instead, stabilizes the vector representation. Hence the so named weak $\mathrm{G_2}$
holonomy of the $\mathrm{SO(7)}$ spin connection $\omega^{AB}$ is the same
thing as the $\mathrm{SO(7)}^+$ holonomy of the  $\mathrm{SO(8)}$ Lie algebra
valued \textit{de Sitter connection} $\left\{\omega^{AB},
e^C\right\}$ introduced in \cite{fermionMassSpectrum} and normally
discussed in the old literature on Kaluza Klein Supergravity.
\par
We have solved Maxwell equation, but we still have to  solve
Einstein equation, namely  (\ref{fresconiB}). To this effect we
note that:
\begin{equation}
  \mathcal{F}_{BIJK}\,\mathcal{F}^{AIJK} \, = \, 24 \, \delta^A_B
  \quad \Longrightarrow \quad \mu = 24
\label{mufixed}
\end{equation}
and we observe that eq.(\ref{fresconiB}) reduces to the following two
conditions on the parameters (see \cite{su3su2u1} for details) :
\begin{eqnarray}
\ft 32 \, \lambda & = & - \left( 24 \, e^2 + \ft 7 2 \, \mu \, f^2 \right)  \nonumber\\
3 \, \nu  & = & 12 \, e^2 + \ft 52 \, \mu \, f^2
\label{twocondos}
\end{eqnarray}
From eq.s (\ref{twocondos}) we conclude that there are only three
possible kind of solutions.
\begin{description}
  \item[a] The flat solutions of type
\begin{equation}
  \mathcal{M}_{11} \, = \, Mink_4 \, \otimes \, \underbrace{\mathcal{M}_7}_{\mbox{Ricci flat}}
\label{minktimespiat}
\end{equation}
where both $D=4$ space-time and the internal $7$-space are Ricci
flat. These compactifications correspond to  $e=0$ and $F_{IJKL} =
0 \, \Rightarrow \, g_{IJKL} = 0$. In this category fall  the
typical twisted tori compactifications which implement the Scherk
and Schwarz mechanism but do not support any internal flux. So, in
view of the results of \cite{chevrolet}, they cannot lead to any
FDA in $D=4$ with minimal subalgebra larger than $\mathbb{G}_7$.
  \item[b] The Freund Rubin solutions of type
\begin{equation}
  \mathcal{M}_{11} \, = \, \mathrm{AdS}_4 \, \otimes \, \underbrace{\mathcal{M}_7}_{\mbox{Einst. manif.}}
\label{adstimesEinst}
\end{equation}
These correspond to anti de Sitter space in $4$-dimensions, whose
radius is fixed by the Freund Rubin parameter $e \ne 0 $ times any
Einstein manifold in $7$--dimensions with no internal flux, namely
$g_{IJKL} = 0$. Once again, in view of \cite{chevrolet}, also
these compactifications lead to FDAs in $D=4$ which have
$\mathbb{G}_7$ as the minimal part.
  \item[c] The Englert type solutions
  \begin{equation}
  \mathcal{M}_{11} \, = \, \mathrm{AdS}_4 \, \otimes \,
  \underbrace{\mathcal{M}_7}_{\begin{array}{c}
    \mbox{Einst. manif.} \\
    \mbox{weak $\mathrm{G_2}$ hol} \
  \end{array}}
\label{adstimeG2}
\end{equation}
These correspond to anti de Sitter space in $4$-dimensions ($e \ne
0$) times a $7$--dimensional Einstein manifold which is
necessarily of weak $\mathrm{G_2}$ holonomy in order to support a
consistent non vanishing internal flux $g_{IJKL}$. These are the
only possible candidate compactifications for the generation of
FDA in $D=4$ in which the  minimal part is a proper extension of
$\mathbb{G}_7$. In the sequel we concentrate on Englert solutions.
\end{description}
In view of the previous discussion we set $\mathcal{F}_{IJKL} \ne 0$
and we complete the analysis of our parameter equations, just
recalling the results presented in \cite{su3su2u1}.
Combining eq.s (\ref{twocondos}) with the previous ones we uniquely
obtain:
\begin{equation}
  \lambda = -30 \, e^2 \quad ; \quad f = \pm \ft 12 \, e
\label{soluzia}
\end{equation}
Summarizing, provided, the manifold $\mathcal{M}_{7}$ has weak $\mathrm{G_2}$ holonomy,
namely provided there exist a weak Killing spinor, satisfying the
condition:
\begin{equation}
  \mathcal{D}_I \, \eta \, = -\, \ft 32 \, e \, \gamma_I \, \eta
\label{weakkispinorBis}
\end{equation}
we obtain a pair of $\mathrm{G_2}$ forms :
\begin{eqnarray}
\Phi & \equiv & \phi_{IJK} \, e^I \wedge e^J \wedge e^K \, = \,
\ft 16 \eta^T \, \gamma_{IJK} \eta \, e^I \wedge e^J \wedge e^K \, \nonumber\\
\Phi^\star & = & \phi^\star_{IJKL}\, e^I \wedge e^J \wedge e^K \wedge e^L
\, = \,  \ft {1}{24} \eta^T \, \gamma_{IJKL} \eta \, e^I \wedge e^J \wedge e^K \wedge e^L
\label{loreipsum}
\end{eqnarray}
satisfying the condition
\begin{equation}
  d\Phi = 12 \, e \, \Phi^\star
\label{intrinsic}
\end{equation}
In this case a unique consistent Englert solution of $M$-theory with internal
fluxes is obtained by setting
\begin{eqnarray}
R^{rs}_{\phantom{ab}tu} & = & -30 \, e^2 \, \delta^{rs}_{tu} \label{AdScurvaBis}\\
\mathcal{R}^{IK}_{JK} & = & 27 \, e^2 \,  \delta^{I}_{J} \label{M7curvaBis}\\
F_{rstu} & = & e \, \epsilon_{rstu} \label{external fluxBis}\\
F_{IJKL} & = & 12 \, e \, \phi^\star_{IJKL}  \label{internalfluxBis}
\end{eqnarray}
It must also be noted that equation (\ref{M7curvaBis}) is not an
independent condition but, according to the previous discussion is a
consequence of eq. (\ref{intrinsic}).
\par
As we already mentioned in the introduction there exist several
compact manifolds of weak $\mathrm{G_2}$ holonomy. In particular
all the coset manifolds $\mathcal{S}/\mathcal{R}$ of weak
$\mathrm{G_2}$ holonomy were classified and studied in the Kaluza
Klein supergravity age
\cite{Awada:1982pk,su3su2u1,D'Auria:1983vy,Castellani:1983tc,classificoleo}
and they were extensively reconsidered in the context of the
AdS/CFT correspondence
\cite{Billo:2000zs,Billo:2000zr,Fre':1999xp,Fabbri:1999hw,Fabbri:1999mk}.
No one so far formulated and answered the question whether there
exist $7$-parameter Lie group manifolds $\mathcal{G}$ of weak
$\mathrm{G_2}$ holonomy. In the next sections we address such a
question and we obtain a negative answer. No such Lie group
manifold exists.

\section{Group theoretical reformulation of weak $\mathrm{G_2}$ holonomy at the spin
connection level} \label{grupporiformulo} Let us address in this
section the question whether there are twisted torii of weak
$\mathrm{G_2}$ holonomy, namely non semisimple group manifolds
with such a property. To this effect a very useful result was
obtained in paper \cite{weakG2} where it was shown that equation
(\ref{intrinsic}) is fully equivalent to a condition imposed
directly on the spin connection of the internal manifold, namely:
\begin{equation}
  \phi^{IJK} \, \omega^{JK} \, = q \,  e^I \quad ; \quad \mbox{with $q = -6\, e$}
\label{alphacondo}
\end{equation}
\par
This reformulation is very useful because can be immediately
translated into a group theoretical language. In order to satisfy eq.
(\ref{alphacondo}) we can just set:
\begin{equation}
  \omega^{IJ} = \omega^{IJ}_{(14)} \, + \, 6 \, q \, \phi^{IJK} \, e^K
\label{6qcondo}
\end{equation}
where $\omega^{IJ}_{(14)}$ is a one--form valued in the
$\mathbf{14}$--dimensional representation of $\mathrm{G_2}$, namely in the
adjoint. Being a one-form it can be expanded along the vielbein and
we can write:
\begin{equation}
  \omega^{IJ}_{(14)} = \overline{\omega}^{IJ}_K \, e^K
\label{omegabar}
\end{equation}
Group theoretically the tensor $\overline{\omega}^{IJ}_K$ is in the
product of the $\mathbf{14}$ with the $\mathbf{7}$ and we have the
decomposition:
\begin{equation}
  \mathbf{14} \times \mathbf{7} = \mathbf{64} \oplus \mathbf{27} \oplus \mathbf{7}
\label{containedreps}
\end{equation}
Hence, a priori, the spin connection of a weak $\mathrm{G_2}$ holonomy
manifold can be parametrized with the above irreducible $\mathrm{G_2}$
tensors. Let us however suppose that our manifold is a group
manifold, characterized by the Maurer Cartan eq.s (\ref{MC1} ) satisfied by
the seven vielbein where $\tau^{I}_{\phantom{I}JK}$ are the structure constants of a
seven dimensional Lie algebra spanned by generators $\left\{ T_J\right\} $ which are dual
to the $1$-forms $e^I$:
\begin{equation}
\left[ T_I \, , \, T_J \right] \, = \, \tau^K_{\phantom{K}IJ} \, T_K
\quad \Leftrightarrow \quad e^I\left( T_J\right)  = \delta^I_J
\label{dualitaMC}
\end{equation}
Let us furthermore suppose that the
above Lie algebra is \textit{volume preserving}, namely satisfies the
additional condition:
\begin{equation}
  \tau^{I}_{\phantom{I}IK} = 0
\label{volumepreserving}
\end{equation}
which was assumed both in the series of papers
\cite{df,ddf,dft1,dft2}, and in
\cite{chevrolet}.
In this case the representation $\mathbf{7}$ is suppressed in the
decomposition (\ref{containedreps}). Indeed, the spin connection,
defined by:
\begin{equation}
  \partial \, e^I \, + \, \omega^{IJ} \, \wedge \, e^J \, = \, 0
\label{de+ome}
\end{equation}
turns out to be related to the structure constants by the simple
formula:
\begin{equation}
  \tau^{I}_{\phantom{I}AB} = \omega^{IB}_A - \omega^{IA}_B
\label{omegatau}
\end{equation}
and equation (\ref{volumepreserving}) simply states that the
representation $\mathbf{7}$ is not present.
\par
If instead we do not make this assumption we still have to add a
seven dimensional representation. In any case by means of this
argument the problem is reduced to its algebraic core. In appendix
\ref{g2formalismo} we recall how the $\mathbf{64}$ and
$\mathbf{27}$ representations are constructed in terms of tensors
satisfying certain $\mathrm{G_2}$ invariant constraints. These
constraints can also be algebraically solved (we did it with a
computer programme in MATHEMATICA) and the corresponding tensors
parametrized by $64$, respectively  $27$ independent parameters,
denoted $\xi^i$ ($i=1,\dots, 64$) and $\alpha^i$ ($i=1,\dots, 64$)
are named $H^{(64)}_{\scriptsize{\begin{array}{cc}
  A & I \\
  B & \null\\
\end{array}}}(\xi)$ and $U_{ABI}^{(27)}(\alpha)$.
In terms of these latter we can  write an ansatz for the spin connection:
\begin{equation}
  \omega^{AB}_I = H^{(64)}_{\scriptsize{\begin{array}{cc}
  A & I \\
  B & \null\\
\end{array}}}(\xi) \, + \, U_{ABI}^{(27)}(\alpha) \,  + \,  6\, q \,
\phi^{ABI}
\label{paramomega}
\end{equation}
which, through the simple formula (\ref{omegatau}), leads to an ansatz
for the corresponding candidate structure constants $\tau^{I}_{\phantom{I}AB}(\xi,\alpha, q)$
depending on the $91$ parameters $(\xi,\alpha)$ plus the parameter $q$. As long as
$q\ne 0$ we are guaranteed to have weak $\mathrm{G_2}$ holonomy. The
problem is that, in order to define a true Lie algebra, the $\tau^{I}_{\phantom{I}AB}(\xi,\alpha, q)$
should satisfy Jacobi identities:
\begin{equation}
  \tau^{I}_{\phantom{I}[AB}(\xi,\alpha, q) \, \tau^{J}_{\phantom{I}C]I}(\xi,\alpha,
  q) \, = \, 0
\label{jacobbus}
\end{equation}
So we should solve the quadratic equations (\ref{jacobbus}) and
find out how many of the $91$ parameters remain free. These,
modulo $\mathrm{G_2}$ rotations, will span the space of
$7$--dimensional Lie algebras of weak $\mathrm{G_2}$ holonomy. At
first sight $91$ seems a quite comfortable number and one  might
expect to find not just one but several solutions. Yet Lie
algebras are tough cookies and  the real case is just the
opposite. No solutions do exist. A direct proof by solving the
quadratic equations (\ref{jacobbus}) is conceivable but certainly
requires a lot of computer time. We can however reach the same
result by focusing on the Ricci tensor defined by the spin
connection (\ref{omegatau}). The Ricci form, if the algebra had
weak $\mathrm{G_2}$ would be proportional to a Kronecker delta and
so, in particular positive definite. In the next section, by
exploiting structural Lie algebra theorems we will be able to
prove that this cannot happen.
\section{The Ricci tensor of metric Lie algebras and the main no-go theorem}
\label{ricciussection} The main object of study in the present
section is the Ricci tensor of a group manifold $\mathcal{G}$,
whose Lie algebra $\mathbb{G}_7$ is characterized by the structure
constants $\tau^{I}_{\phantom{I}AB}$, alternatively defined by the
commutation relations (\ref{dualitaMC}) or by the Maurer Cartan
equations (\ref{MC1}). The metric of the manifold is given by:
\begin{equation}
  ds^2_{\mathcal{G}} = \eta_{IJ} \, e^I \, \otimes \, e^J
\label{metricaG}
\end{equation}
where $\eta_{IJ}$ is the flat metric. In agreement with the notations of \cite{su3su2u1}, which
we have adopted throughout our entire discussion, the spin connection $1$-form
is defined by the vanishing torsion equation:
\begin{equation}
  de^I - \, \omega^{IJ} \, \wedge \, e^K \, \eta_{JK} \, = \, 0
\label{vanatorsio}
\end{equation}
and the Riemann $2$-form is normalized as follows:
\begin{equation}
\mathcal{R}^{AB} \, \equiv \,   d\omega^{AB} \, - \, \omega^{AI} \, \wedge \, \omega^{JB} \,
  \eta_{IJ} \, = \, \mathcal{R}^{AB}_{\phantom{AB}PQ} \, e^P \,
  \wedge \, e^Q
\label{Riemanntensore}
\end{equation}
so that the Ricci tensor is finally defined by:
\begin{equation}
  \mathrm{Ric}[\tau]^I_{\phantom{I}J} \, \equiv \, \mathcal{R}^{IM}_{\phantom{AB}JM}
\label{riccidefi}
\end{equation}
By explicit calculation, from the above formulae
(\ref{vanatorsio}-\ref{riccidefi}) one obtains
\begin{eqnarray}
\mathrm{Ric}[\tau]^I_{\phantom{I}J} & = & \frac{1}{4} \left(
\eta^{IN} \, \tau^{K}_{\phantom{K}NS} \,
\tau^{S}_{\phantom{S}JK} \, + \,
\eta^{IQ} \, \eta^{SM} \, \eta_{PN} \,
\tau^{P}_{\phantom{P}QS} \,
\tau^{N}_{\phantom{N}JM} \,- \, \frac{1}{2}\,
\eta^{PQ} \, \eta^{SM} \, \eta_{JN} \,
\tau^{I}_{\phantom{I}PS} \,
\tau^{N}_{\phantom{S}QM}\,
\right.
 \nonumber\\
 \null & \null & \left. + \, \eta^{NK} \,
 \tau^{I}_{\phantom{I}NJ} \,
\tau^{S}_{\phantom{S}KS} \, - \,
\, \eta^{IN} \,
 \tau^{K}_{\phantom{I}NJ} \,
\tau^{S}_{\phantom{S}KS}
\, + \,
\eta^{PQ} \, \eta^{IM} \, \eta_{JN} \,
\tau^{S}_{\phantom{S}PS} \,
\tau^{N}_{\phantom{N}QM}\, \right )
\label{AlRiccitens}
\end{eqnarray}
where we have included also the trace terms of type
$\tau^{S}_{\phantom{S}PS}$, which means that we have relaxed the area
preserving condition (\ref{volumepreserving}). In our case of
interest, namely in compactifications of $M$-theory the flat metric $\eta^{IJ}$ is chosen to be the negative
Euclidean metric:
\begin{equation}
  \eta^{IJ} \, = \, - \, \delta^{IJ}
\label{minusdelta}
\end{equation}
This corresponds to the conventions of \cite{su3su2u1} and follows
from the fact that the $11$--dimensional signature was chosen mostly
minus. When eq.(\ref{minusdelta}) is adopted
eq. (\ref{AlRiccitens}) can be rewritten in a very convenient matrix form:
\begin{eqnarray}
\mathrm{Ric}[\tau]^I{}_{J} & = & -\ft 14 \, \mbox{Tr}\left[ \tau_I
\,\tau_J \right ] \, - \, \ft 14 \mbox{Tr}\left[ \tau_I \,
\tau_J^T\right] \, + \ft 18 \, \sum_{K=1}^7 \left(\tau_K \,
\tau_K^T \right)^{IJ}  \nonumber\\
\null & \null & \, - \, \ft 14 \,
\sum_{K=1}^7 \, \tau^{I}_{\phantom{I}KJ}  \,
\mbox{Tr}\left[
\tau_K \right ] \, + \,
 \ft 14 \,
\sum_{K=1}^7 \, \tau^{K}_{\phantom{I}IJ}  \,
\mbox{Tr}\left[
\tau_K \right ] \,  - \,
 \ft 14 \,
\sum_{K=1}^7 \, \tau^{J}_{\phantom{I}KI}  \,
\mbox{Tr}\left[
\tau_K \right ]
\label{Riccimatrona}
\end{eqnarray}
where $(\tau_I)^A_{B} \equiv \tau^A_{\phantom{A}IB}$ is the matrix
representing the generator $T_I$ in the adjoint representation and
$\sum_{K=1}^7 \left(\tau_K \, \tau_K^T \right)^{IJ}$ denotes the
matrix $\sum_{K,L=1}^7 \,\tau^I{}_{KL} \, \tau^J{}_{KL}$.
\par
As we have extensively discussed in previous sections,  the problem
of establishing whether or not, there exist \textit{bona
fide} compactifications of M-theory on twisted tori with non trivial fluxes has been
reduced, in the absence of sources, to the question whether or not,
there exist $7$--dimensional Lie algebras of weak $\mathrm{G_2}$ holonomy.
Relying on the general representation of the Ricci tensor provided by
eq.(\ref{Riccimatrona}) we can now state our main negative result as
the following:
\begin{teorema}\label{mainoteoremo}
There exists no real $7$-dimensional metric Lie algebra $\mathbb{G}_7$ of weak $\mathrm{G_2}$
holonomy
\end{teorema}
\begin{proofteo}
{\rm We prove this theorem in a series of steps by exhaustion of all
possible cases. First let us explain the general strategy of the proof.
Given a real Lie algebra $\mathbb{G}_7$ identified, in any given basis of
generators $\left\{ T_I \right\} $, by its structure constants
$\tau^{I}_{JK}$, normalized as in eq.(\ref{dualitaMC}), a metric ${\mathbf{g}}$ on
$\mathbb{G}_7$ is a bilinear, symmetric, non degenerate, negative definite $2$-form:
\begin{eqnarray}
< \, , \, > _\mathbf{g} \quad : \quad \mathbb{G} \, \otimes \, \mathbb{G}
\,\rightarrow \, \mathbb{R}
\label{gforma}
\end{eqnarray}
As usual $< \, , \, > _\mathbf{g}$ is determined by giving its values in
a basis, namely on the generators $\left\{ T_I \right\} $:
\begin{equation}
  < T_I\, , \, T_J > _\mathbf{g} \, = \, \mathbf{g}_{IJ}
\label{orgaMetric}
\end{equation}
By hypothesis the matrix $\mathbf{g}_{IJ}$ is symmetric ($\mathbf{g}_{IJ} =
\mathbf{g}_{JI}$), non-degenerate ($\mathrm{det} \,\mathbf{ g}\,  \ne \,
0$) and negative definite, namely all of its eigenvalues are strictly
negative ($\lambda_I < 0 $). The holonomy, in particular weak $\mathrm{G_2}$ holonomy,
is a property of the metric, so that on the same algebra $\mathbb{G}_7$
there can be a metric $\mathbf{g}_{IJ}$ with a certain holonomy and
another one ${\widetilde{\mathbf{g}}}_{IJ}$ with a different
holonomy. The crucial observation, however, is that, once the
signature of $\mathbf{g}_{IJ}$ has been fixed (in our case the
euclidean negative one), the choice of the metric is equivalent to a
change of basis in the algebra. Indeed under the latter $T_I \, \rightarrow \, \mathcal{A}_I^{\phantom{I}J} \, T_J$,
where $\mathrm{det} \mathcal{A} \ne 0$  is a non degenerate real matrix, the metric changes as follows:
\begin{equation}
  \mathbf{g} \, \rightarrow \, \widetilde{\mathbf{g}} \, = \, \mathcal{A}^T \,
  \mathbf{g} \, \mathcal{A}
\label{changiometra}
\end{equation}
and by means of a suitable real $\mathcal{A}$ we can always obtain
$\widetilde{\mathbf{g}}_{IJ}= - \delta_{IJ}$. As it is evident from
its definition (\ref{Riccimatrona}) also the Ricci form changes in
the same way as the metric:
\begin{equation}
  \mathbf{Ric}[\tau] \, \rightarrow \, {\mathbf{Ric}}[\widetilde{\tau}] \, = \, \mathcal{A}^T \,
  \mathbf{Ric}[\tau] \, \mathcal{A}
\label{changioriccio}
\end{equation}
So that, instead of considering the space of metrics on each given
$7$--dimensional Lie algebra $\mathbb{G}_7$, it is sufficient,
fixing the metric to be the standard one, ($i.e.$
${\mathbf{g}}_{IJ}= - \delta_{IJ}$), to consider all possible Lie
algebra bases, parametrized by the elements  $\mathcal{A} \, \in
\, \mathrm{GL(7,\mathbb{R})}$ of the general linear group in
$7$--dimensions. As we have recalled in previous sections, if the
spin connection of a $7$--manifold $\mathcal{M}_7$ has weak
$\mathrm{G_2}$--holonomy, namely if $\omega^{IJ}$ satisfies the
defining condition (\ref{alphacondo}), then the intrinsic Ricci
tensor ${\mathbf{Ric}}^I{}_{J} \equiv
\mathcal{R}^{IM}_{\phantom{IM}JM}$ is automatically proportional
to $\delta^I{}_{J}$ with a positive coefficient, namely it is a
positive definite symmetric $2$-form. This observation provides a
very severe necessary (although  not yet sufficient condition) for
weak $\mathrm{G_2}$ holonomy: the Ricci form   of the candidate
$\mathbb{G}_7$ Lie algebra, equipped with a negative definite
metric $\mathbf{g} < 0$ should instead be positive definite
$\mathbf{Ric}
>0$. The value of this criterion is that it depends only on the
choice of the algebra and not on the choice of the basis, or
equivalently of the metric. So if we are able to prove that for all
$7$-dimensional Lie algebras the Ricci form is never positive
definite, then we have proved our theorem.
\par
As we anticipated, we proceed by exhaustion of all the possible
cases, relying on the fundamental structural theorem by Levi which states
that
\begin{lemma}\label{Levitheorem}
Any Lie algebra $\mathbb{G}$ of dimension ${\mathrm{dim}}\, \mathbb{G} = n$ is the semidirect product of a
semisimple Lie algebra $\mathbb{L}(\mathbb{G})$ of dimension ${\mathrm{dim}}\, \mathbb{L}(\mathbb{G})\, = \, m$,
called the Levi subalgebra $\mathbb{L}(\mathbb{G}) \subset \mathbb{G}$ with a solvable
ideal $\mathrm{Rad}(\mathbb{G}) \subset \mathbb{G} $ of dimension ${\mathrm{dim}}\, \mathrm{Rad}(\mathbb{G})\, =
\, q$ so that $n=m+q$. The solvable
ideal $\mathrm{Rad}(\mathbb{G})$ is named the radical of $\mathbb{G}$
\end{lemma}
For the proof of Levi's theorem we refer the reader to standard
textbooks as \cite{jacobson} or \cite{varadarajan}. Applying  it to
our case we can just go by dimension of the radical, namely by values of $q$.
\begin{description}
  \item[($q=0$)] This corresponds to the case when $\mathrm{Rad}(\mathbb{G}_7) =
  0$, namely when $\mathbb{G}_7$ is semisimple. Yet there are no semisimple Lie
  algebras of dimension $7$, so this case is already ruled out.
  \item[($q=7$)] This is the extreme opposite case when the Levi
  subalgebra vanishes $\mathbb{L}(\mathbb{G}_7)=0$, namely $\mathbb{G}_7$ is
  completely solvable. By definition this means that the first
  derivative of the algebra $\mathcal{D}\mathbb{G}_7 \equiv \left[\mathbb{ G}_7 \, , \, \mathbb{G}_7
  \right]$ is necessarily a proper subspace $i.e.$ $\mathrm{dim} \, \mathcal{D}\mathbb{G}_7 \,  <
  7$. Henceforth we can write the following orthogonal decomposition
  of $\mathbb{G}_7$:
\begin{equation}
\mathbb{G}_7 =\mathbb{K}_0 \, \oplus \, \mathcal{D}\mathbb{G}_7
\label{ortoDG+K}
\end{equation}
where $\mathrm{dim} \mathbb{K}_0 \ge 1$. Let us accordingly subdivide
the index range as it follows $I = \left\{i,\alpha\right\}$ where $i=,1 \dots, {\mathrm{dim}
\, \mathbb{K}_0}$ spans the  $0$-grading subspace while $\alpha =  {\mathrm{dim}
\, \mathbb{K}_0}+1 \, , \dots, 7$ spans the ideal
$\mathcal{D}\mathbb{G}_7$. In these notations we have that the
structure constants $\tau^{i}_{\phantom{i}IJ} =0$ vanish for all
values of the lower indices $IJ$. Considering next
eq.(\ref{Riccimatrona}) we can calculate the entries $ij$ of the
Ricci form. We immediately find:
\begin{equation}
  \mathbf{Ric}^i{}_{ j} = - \frac{1}{2} \, \mathrm{Tr} \, \left (\tau_i ^{(S)} \, \tau_j ^{(S)}
  \right )\,
\label{squirro}
\end{equation}
where by $\tau_i ^{(S)}$ we have denoted the symmetric part of the
adjoint matrix $\tau_{\phantom{P}iQ}^P$ representing the generator
$T_i \in \mathbb{K}_0$. That eq.(\ref{squirro}) is correct is easily
understood by means of the following argument. The first two terms in
the r.h.s of eq. (\ref{Riccimatrona}) combine into the r.h.s of
(\ref{squirro}), all the other terms vanish since the index $i$ or
$j$ appears as choice of the upper index of
$\tau^{I}_{\phantom{I}JK}$. Hence if calculate the Ricci norm of a
vector lying in the grading $0$-subspace $X=c^i \, T_i$ we find that
it is strictly non positive:
\begin{equation}
\forall X \, \in \, \mathbb{K}_0 \quad : \quad
\mathbf{ Ric}(X,X) = -\frac{1}{4} \, \mathrm{Tr} \left( \tau^{(S)}_X \right)
  ^2 \, < \, 0
\label{horror1}
\end{equation}
This result suffices to prove that $\mathbf{Ric}$ cannot be a
positive definite $2$--form and to rule out also this case.
  \item[($q=2$)] Is ruled out because there are no semisimple Lie
  algebras of dimension 5 and hence the Levi subalgebra cannot exist
  for $m=7-2$
  \item[($q=3$)] Is ruled out because there are no semisimple Lie
  algebras of dimension 4 and hence the Levi subalgebra cannot exist
  for $m=7-3$
  \item[($q=5$)] Is ruled out because there are no semisimple Lie
  algebras of dimension 2 and hence the Levi subalgebra cannot exist
  for $m=7-5$
  \item[($q=4$)] In this case the Levi subalgebra
  $\mathbb{L}(\mathbb{G}_7)$ has dimension $m=3$ and there are just
  two possible real simple algebras of that dimensions namely, either
  $\mathbb{L}(\mathbb{G}_7)=\mathrm{SO(3)}$ or
  $\mathbb{L}(\mathbb{G}_7)=\mathrm{SL(2,\mathbb{R})}$.
In appendix \ref{Bappendo} we classify by exhaustion all the seven
dimensional algebras with Levi subalgebra either $\mathrm{SO(3)}$
(in subsection \ref{leviso3} and relative subsubsections) or $\mathrm{SL(2,\mathbb{R})} \sim \mathrm{SO(1,2)}$
(in subsection \ref{Sl2casi} and relative subsubsections). For each
of these algebras we calculate the Ricci form and we show that in
every case it has indefinite signature (both positive and non
positive eigenvalues) so that all cases are excluded. A general
observation is that compact generators lead to non negative
eigenvalues of the Ricci form
while non compact semisimple or nilpotent generators lead to non
positive ones.
\item[($q=1$)] In this case the Levi subalgebra is either of the
following three cases:
\begin{equation}
 \mathbb{ L}(\mathbb{G}_7) = \cases{\mathrm{SO(3)} \oplus \mathrm{SO(3)} \sim \mathrm{SO(4)}\cr
  \mathrm{SO(3)} \oplus \mathrm{SO(1,2)} \cr
\mathrm{SO(1,2)} \oplus \mathrm{SO(1,2)} \sim \mathrm{SO(2,2)}\cr}
\label{trecasoni}
\end{equation}
and the algebra is just the direct sum of its Levi subalgebra with
its $1$-dimensional radical since there are no $1$-dimensional
representations of $\mathbb{ L}(\mathbb{G}_7) $ except the
singlet. Hence the Ricci form is block-diagonal $3+3+1$ and the last
generator, the singlet contributes a zero eigenvalue. Hence also this
last case is excluded.
\end{description}
}
This concludes the proof of our non go theorem.
$\blacksquare$
\end{proofteo}
Before ending the present section we present here a second formal
proof of the main Theorem. The Ricci tensor ${\bf Ric}^I{}_J$ in
our case is a symmetric $7\times 7$ matrix. If it were positive
definite, then, for any non--vanishing seven dimensional vector
${\bf V}$ we would have:
\begin{eqnarray}
{\bf V}^T\,{\bf Ric}\,{\bf V}&>&0
\end{eqnarray}
Below we show that there always exists a ${\bf V}\neq 0$ so that
${\bf V}^T\,{\bf Ric}\,{\bf V}\le 0$. This implies in turn that
${\bf Ric}^I{}_J$ cannot be positive definite and proves Theorem
\ref{mainoteoremo}. Let us start considering the case in which the
structure constants $\tau$ are traceless $\tau^I{}_{JI}=0$, which
is relevant to the compactifications on  twisted--tori studied in
the literature.
\paragraph{$\tau^I{}_{JI}=0$ case:} Let us decompose ${\rm
Rad}(\mathbb{G}_7)$ as follows:
\begin{eqnarray}
{\rm Rad}(\mathbb{G}_7)&=&\mathbb{K}_0+\mathcal{D}(\mathbb{G}_7)
\end{eqnarray}
and label by $u,v=1,\dots, {\rm dim}(\mathbb{L}(\mathbb{G}_7))$
the generators of the Levi subalgebra, by $i,j\dots$ the
generators of $\mathbb{K}_0$ and by $p,q\dots$ the generators of
${\rm Rad}(\mathbb{G}_7)$. Then the only non--vanishing entries of
$\tau^i{}_{JK}$ are $\tau^i{}_{up}$ and  a basis of
$\mathbb{L}(\mathbb{G}_7)$  can be chosen for which the following
properties hold:
\begin{eqnarray}
\tau^p{}_{uq}\tau^p{}_{vq}&=&\tau^q{}_{up}\tau^q{}_{vp}\label{props}
\end{eqnarray}
Let us compute the entries ${\bf Ric}^i{}_j$. The (positive)
contribution from the third term of (\ref{Riccimatrona}), namely
$(1/4)\,\sum_{u,p=1}^7\,\tau^i{}_{up}\,\tau^j{}_{up}$ is cancelled
by an opposite contribution $-\ft 1 4
\sum_{u,p=1}^7\,\tau^p{}_{ui}\,\tau^p{}_{uj}$ coming from the
first two terms, in virtue of  eq. (\ref{props}). We are left with
the following matrix:
\begin{eqnarray}
{\bf
Ric}^i{}_j&=&-\frac{1}{4}\sum_{p,q=1}^7\,\tau^p{}_{iq}\,(\tau^p{}_{jq}+\tau^q{}_{jp})\le
0
\end{eqnarray}
Therefore, if we choose the following vector ${\bf V}$:
\begin{eqnarray}
{\bf V}&=&(0,\dots, 0, V_i, 0,\dots, 0)\label{v}
\end{eqnarray}
we have ${\bf V}^T\,{\bf Ric}\,{\bf V}\le 0$ and the Theorem is
proven.
\paragraph{$\tau^I{}_{JI}\neq 0$ case:} If
$\mathbb{L}(\mathbb{G}_7)$ is compact, the $\tau_u$ matrices are
antisymmetric and ${\rm Tr}(\tau_u)=0$. Moreover, since the
commutator of two elements of $\mathbb{K}$ expands in nilpotent
generators belonging to $\mathcal{D}(\mathbb{G}_7)$, whose trace
therefore vanishes, we also have $\tau^K{}_{ij}\,{\rm
Tr}(\tau_K)=0$. In this case the terms in ${\bf Ric}^i{}_j$
depending on ${\rm Tr}(\tau_I)$ do not contribute and thus we can
choose the same seven--vector ${\bf V}$ as in (\ref{v}) to prove
that ${\bf Ric}$ is not positive definite.\par In the case in
which $\mathbb{L}(\mathbb{G}_7)$ is a non--compact semisimple
algebra, let us label by $u_a$ and $u_s$ the compact and
non--compact generators respectively. We can have ${\rm
Tr}(\tau_{u_s})\neq 0$ and the previous proof would not hold. Let
us consider ${\bf Ric}^{u_s}{}_{v_s}$ instead. The only non
vanishing entries of $\tau$ having one index $u_s$ are:
\begin{eqnarray}
\tau^{u_a}{}_{u_s v_s}=\tau^{v_s}{}_{u_s
u_a}\,\,;\,\,\,\tau^{p}{}_{u_s q}=\tau^{q}{}_{u_s p}
\end{eqnarray}
 The contribution from the last three terms of
 (\ref{Riccimatrona}) vanishes since it involves only ${\rm
 Tr}(\tau_{u_a})$. The only positive term (the third) in (\ref{Riccimatrona})
 reads $\ft 1 4\, \sum_{u_a w_s=1}^7\, \tau^{u_s}{}_{u_a w_s}\,\tau^{v_s}{}_{u_a
 w_s}$ and it is cancelled by an opposite contribution from the first two
 terms. The remaining matrix is non--positive: ${\bf
Ric}^{u_s}{}_{v_s}\le 0$ since it has contributions only from the
first two terms and thus it suffices to choose the seven--vector
${\bf V}$ with non--vanishing entries only along the non--compact
generators of the Levi subalgebra to have ${\bf V}^T\,{\bf
Ric}\,{\bf V}\le 0$.
\section{Conclusions}
\label{concludo} In this paper we have proved that no Lie group
$7$-manifolds of weak $\mathrm{G_2}$ holonomy do exist. We have
also shown that the notion of weak $\mathrm{G_2}$ holonomy is
identical with the notion of Englert $7$-manifolds introduced and
used in the literature on Kaluza Klein supergravity two decades
ago. Our analysis rules out the possibility of introducing on a
bosonic background a non trivial 4--form flux, and thus, in the
light of the results obtained in \cite{chevrolet}, it follows that
on these solutions the minimal part of the FDA coincides with the
algebra $\mathbb{G}_7$ gauged by the Kaluza--Klein vectors. The
triviality of the 4--form flux on bosonic backgrounds was also
found in \cite{dft2,dft4}. As a byproduct of our analysis we rule
out also (unwarped, supersymmetric and non--supersymmetric)
anti--de Sitter four dimensional vacua (Freund--Rubin) from
$M$--theory compactifications on the so called ``twisted'' tori.
Yet, as we have spelled out in the introduction, this no go
theorem applies only to the case of the field equations of
$M$--theory in the bulk, with no source terms (no M2 or M5 branes)
and no warp factor. It is then obvious that the hunting ground is
singled out by our analysis. One should
\begin{description}
  \item[a] Repeat the cohomological analysis of papers
\cite{df,ddf,dft1,dft2} and \cite{chevrolet} in the presence of a
warp factor.
  \item[b] Analyze the extent to which a warp factor can cope for the
  absence of a Lie algebra of weak $\mathrm{G_2}$ holonomy
  \item[c] Analyze the extent to which the $M2$ and $M5$ brane
  contributions to the stress energy tensor can help in introducing
  consistent internal fluxes.
\end{description}
These are the guidelines for any further investigation.
\par
It is also worth mentioning that a similar analysis for twisted tori
compactifications of Type IIB theory is still missing and should
certainly be considered.

\section{Acknowledgements}
M.T. would like to thank R. D'Auria for useful clarifications.\par
 Work supported by the European
Community's Human Potential Program under contract
MRTN-CT-2004-005104 `Constituents, fundamental forces and
symmetries of the universe'.

\newpage
\appendix
\section{$\mathrm{G_2}$ formalism: notations and normalizations}
\label{g2formalismo}
The group $\mathrm{G_2}$ is defined as the stability subgroup of an $\mathrm{SO(7)}$
spinor so that we have the branching rule:
\begin{equation}
  \mathbf{8} \stackrel{G_2}{\longrightarrow} \mathbf{7} \, \oplus \, \mathbf{1}
\label{8in7p1}
\end{equation}
Correspondingly  the fundamental vector representation of
$\mathrm{SO(7)}$:
\begin{equation}
  \mathbf{7} \stackrel{G_2}{\longrightarrow} \mathbf{7}
\label{7in7}
\end{equation}
remains instead irreducible. For this reason we begin by constructing a set of gamma matrices satisfying the $\mathrm{SO(7)}$
Clifford algebra in the form:
\begin{equation}
  \left \{ \gamma_I \, , \, \gamma_J \right\} = - 2\, \delta_{IJ}
\label{gammadefi}
\end{equation}
The minus sign in eq.(\ref{gammadefi}) is due to our choice of a
mostly minus metric in the conventions for $M$--theory, so that
the internal $7$--dimensions have a negative Euclidean metric.
This has special advantages. Indeed with the choice
(\ref{gammadefi}) the gamma matrices are all real and
antisymmetric. Indeed an explicit realization of this Clifford
algebra, given by real antisymmetric matrices:
\begin{equation}
  \gamma_I^\star = \gamma_I \quad ; \quad \gamma_I^T = - \gamma_I
\label{realantisym}
\end{equation}
is given below:
{\scriptsize
\begin{equation}
  \begin{array}{ccccccc}
    \gamma_1 & = & \left( \matrix{
      0 & 0 & 0 & 0 & 0 & 0 & 1 & 0 \cr 0 & 0 & 1 & 0 & 0 & 0 & 0 & 0 \cr 0 & \
- 1 & 0 & 0 & 0 & 0 & 0 & 0 \cr 0 & 0 & 0 & 0 & 0 & 0 & 0 & 1 \cr 0 & 0 & 0 & \
0 & 0 & - 1 & 0 & 0 \cr 0 & 0 & 0 & 0 & 1 & 0 & 0 & 0 \cr -
                                        1 & 0 & 0 & 0 & 0 & 0 & 0 & 0 \cr 0 & \
0 & 0 & - 1 & 0 & 0 & 0 & 0 \cr  } \right)  & ; & \gamma_2 & = & \left( \matrix{ 0 & 1 & 0 & 0 & 0 & 0 & 0 & 0 \cr -
                                        1 & 0 & 0 & 0 & 0 & 0 & 0 & 0 \cr 0 & \
0 & 0 & 0 & 0 & 0 & 1 & 0 \cr 0 & 0 & 0 & 0 & -
                                        1 & 0 & 0 & 0 \cr 0 & 0 & 0 & 1 & 0 & \
0 & 0 & 0 \cr 0 & 0 & 0 & 0 & 0 & 0 & 0 & 1 \cr 0 & 0 & -
                                        1 & 0 & 0 & 0 & 0 & 0 \cr 0 & 0 & 0 & \
0 & 0 & - 1 & 0 & 0 \cr  }\right)  \\
\null & \null & \null & \null & \null & \null & \null \\
    \gamma_3 & = &\left(  \matrix{ 0 & 0 & 0 & 0 & 0 & 0 & 0 & - 1 \cr 0 & 0 & 0 & 0 & 0 & -
                                        1 & 0 & 0 \cr 0 & 0 & 0 & 0 & 1 & 0 & \
0 & 0 \cr 0 & 0 & 0 & 0 & 0 & 0 & 1 & 0 \cr 0 & 0 & -
                                        1 & 0 & 0 & 0 & 0 & 0 \cr 0 & 1 & 0 & \
0 & 0 & 0 & 0 & 0 \cr 0 & 0 & 0 & -
                                        1 & 0 & 0 & 0 & 0 \cr 1 & 0 & 0 & 0 & \
0 & 0 & 0 & 0 \cr  } \right)  & ; & \gamma_4 & = &\left(  \matrix{ 0 & 0 & 0 & 0 & 1 & 0 & 0 & 0 \cr 0 & 0 & 0 & -
                                        1 & 0 & 0 & 0 & 0 \cr 0 & 0 & 0 & 0 & \
0 & 0 & 0 & 1 \cr 0 & 1 & 0 & 0 & 0 & 0 & 0 & 0 \cr -
                                        1 & 0 & 0 & 0 & 0 & 0 & 0 & 0 \cr 0 & \
0 & 0 & 0 & 0 & 0 & - 1 & 0 \cr 0 & 0 & 0 & 0 & 0 & 1 & 0 & 0 \cr 0 & 0 & -
                          1 & 0 & 0 & 0 & 0 & 0 \cr  } \right)  \\
                          \null & \null & \null & \null & \null & \null & \null \\
    \gamma_5 & = &\left(  \matrix{ 0 & 0 & 0 & 1 & 0 & 0 & 0 & 0 \cr 0 & 0 & 0 & 0 & 1 & 0 & 0 & 0 \cr \
0 & 0 & 0 & 0 & 0 & 1 & 0 & 0 \cr - 1 & 0 & 0 & 0 & 0 & 0 & 0 & 0 \cr 0 & -
                                        1 & 0 & 0 & 0 & 0 & 0 & 0 \cr 0 & 0 & \
- 1 & 0 & 0 & 0 & 0 & 0 \cr 0 & 0 & 0 & 0 & 0 & 0 & 0 & -
                                  1 \cr 0 & 0 & 0 & 0 & 0 & 0 & 1 & 0 \cr  } \right)  & ; & \gamma_6 & = &\left(  \matrix{ 0 & 0 & 0 & 0 & 0 & -
                                        1 & 0 & 0 \cr 0 & 0 & 0 & 0 & 0 & 0 & \
0 & 1 \cr 0 & 0 & 0 & 1 & 0 & 0 & 0 & 0 \cr 0 & 0 & -
                                        1 & 0 & 0 & 0 & 0 & 0 \cr 0 & 0 & 0 & \
0 & 0 & 0 & -
                                        1 & 0 \cr 1 & 0 & 0 & 0 & 0 & 0 & 0 & \
0 \cr 0 & 0 & 0 & 0 & 1 & 0 & 0 & 0 \cr 0 & -
                              1 & 0 & 0 & 0 & 0 & 0 & 0 \cr  } \right)  \\
                              \null & \null & \null & \null & \null & \null & \null \\
    \gamma_7 & = &\left(  \matrix{ 0 & 0 & -
                                        1 & 0 & 0 & 0 & 0 & 0 \cr 0 & 0 & 0 & \
0 & 0 & 0 & 1 & 0 \cr 1 & 0 & 0 & 0 & 0 & 0 & 0 & 0 \cr 0 & 0 & 0 & 0 & 0 & 1 \
& 0 & 0 \cr 0 & 0 & 0 & 0 & 0 & 0 & 0 & 1 \cr 0 & 0 & 0 & -
                                        1 & 0 & 0 & 0 & 0 \cr 0 & -
                                        1 & 0 & 0 & 0 & 0 & 0 & 0 \cr 0 & 0 & \
0 & 0 & - 1 & 0 & 0 & 0 \cr  }\right)  & \null & \null & \null & \null \
  \end{array}
\label{gammas}
\end{equation}
}
Let $e^I$ be the vielbein of the considered $7$-dimensional space.
As it is well known we can construct a $\mathrm{G_2}$-invariant three-form (defining the $\mathrm{G_2}$--structure
of the corresponding $7$-manifold) by writing:
\begin{equation}
  \Phi = \frac{1}{6} \, \eta^T \, \gamma_{IJK} \, \eta \, e^I \wedge
  e^J \wedge e^K \equiv \phi_{IJK} \, \eta \, e^I \wedge
  e^J \wedge e^K
\label{Phiform}
\end{equation}
where $\eta$ is the $\mathrm{G_2}$ invariant $8$-component commuting spinor which, by choice of normalization,
we take to be the following one:
\begin{equation}
  \eta = \left (   \begin{array}{c}
    0 \\
    0 \\
    0 \\
    0 \\
    0 \\
    0 \\
    0 \\
    1 \
  \end{array}\right )
\label{etaspinor}
\end{equation}
With this choice we can calculate the explicit expression of the $\mathrm{G_2}$ three-form
and we obtain:
\begin{equation}
\Phi =  e^{1}\wedge  e^{2}\wedge  e^{7} + e^{1}\wedge  e^{3}\wedge  e^{5} -
  e^{1}\wedge  e^{4}\wedge  e^{6} - e^{2}\wedge  e^{3}\wedge  e^{6} -
  e^{2}\wedge  e^{4}\wedge  e^{5} + e^{3}\wedge  e^{4}\wedge  e^{7} +
  e^{5}\wedge  e^{6}\wedge  e^{7}
\label{G2_3form}
\end{equation}
The reader can observe that we have calibrated the form of our gamma
matrices (\ref{gammadefi}) in such a way that the expression for
$\Phi$ coincides with that usually adopted in the mathematical
literature on $G$-structures (see for instance \cite{salamone}).
The dual $\mathrm{G_2}$ invariant four--form is defined by:
\begin{equation}
  \Phi_\star \equiv \frac{1}{24} \, \eta^T \, \gamma_{IJKL}\, \eta \,
  e^I \wedge e^J \wedge e^K \wedge e^L \, \equiv \, \phi^\star_{IJKL}\,
  e^I \wedge e^J \wedge e^K \wedge e^L
\label{Phistardefi}
\end{equation}
and its explicit expression is the following:
\begin{eqnarray}
\Phi_\star & = & - e^{1}\wedge  e^{2}\wedge  e^{3}\wedge  e^{4} -
  e^{1}\wedge  e^{2}\wedge  e^{5}\wedge  e^{6} -
  e^{1}\wedge  e^{3}\wedge  e^{6}\wedge  e^{7} -
  e^{1}\wedge  e^{4}\wedge  e^{5}\wedge  e^{7} \nonumber\\
\null & \null & -
  e^{2}\wedge  e^{3}\wedge  e^{5}\wedge  e^{7} +
  e^{2}\wedge  e^{4}\wedge  e^{6}\wedge  e^{7} -
  e^{3}\wedge  e^{4}\wedge  e^{5}\wedge  e^{6}
\label{phistar}
\end{eqnarray}
The components of these invariant $G_2$ forms are dual to each other
since they satisfy the relation:
\begin{equation}
  - \frac{1}{6} \, \epsilon_{IJKABCD} \, \phi^{\star}_{ABCD} \, = \,
  \phi_{IJK}
\label{dualphiphistar}
\end{equation}
Next we introduce the projection operators onto the $\mathbf{14}$ and the $\mathbf{7}$
dimensional representations of $\mathrm{G_2}$. We recall that the adjoint of
$\mathrm{SO(7)}$ splits in the following way with respect to the
$\mathrm{G_2}$ subalgebra:
\begin{equation}
  \mathbf{21} \, \stackrel{\mathrm{G_2}}{\longrightarrow} \, \underbrace{\mathbf{14}}_{adjoint} \, \oplus \,
  \underbrace{\mathbf{7}}_{fundamental}
\label{21splitta}
\end{equation}
The adjoint of $\mathrm{SO(7)}$ is given by an antisymmetric tensor $M^{IJ} =
- M^{JI}$ whose decomposition is as follows:
\begin{equation}
  M^{IJ} = M^{IJ}_{14} + M^{IJ}_7
\label{lobotomia}
\end{equation}
The $\mathbf{14}$ part is defined by the $\mathrm{G_2}$ invariant condition:
\begin{equation}
  \phi^{IJK} \, M^{JK}_{14} \, = \, 0
\label{14condo}
\end{equation}
Correspondingly the projector onto the $\mathbf{14}$ part of any antisymmetric
tensor is given by:
\begin{equation}
  P_{(14)\phantom{IJ}KL}^{IJ} = \frac{2}{3} \, \delta^{IJ}_{KL} \, + \,
  4 \, \phi^{IJKL}_\star
\label{P14proj}
\end{equation}
while the projector onto the $\mathbf{7}$ part is given by:
\begin{equation}
  P_{(7)\phantom{IJ}KL}^{IJ} = \frac{1}{3} \, \delta^{IJ}_{KL} \, - \,
  4 \, \phi^{IJKL}_\star
\label{P7proj}
\end{equation}
Applied onto the two representations the dual form
$\phi_\star^{IJKL}$ is a diagonal operator with two different
eigenvalues:
\begin{eqnarray}
\phi_\star^{IJKL} \, M^{KL}_{14} & =  & \frac{1}{12} \, M^{IJ}_{14} \nonumber\\
\phi_\star^{IJKL} \, M^{KL}_{7} & =  & - \frac{1}{6} \, M^{IJ}_{7}
\label{ordolaro}
\end{eqnarray}
The expression of these projection operators and the calculation of the eigenvalues (\ref{ordolaro}) was presented long time
ago in \cite{fermionMassSpectrum}. More recently appeared often in
the literature on flux compactifications and specifically in
\cite{weakG2}.
\subsection{A Fierz identity and the $\mathrm{G_2}$ decomposition of the $\mathbf{35}$ representation of $\mathrm{SO(7)}$}
An important relation is provided by the following identity (see for instance \cite{weakG2}):
\begin{eqnarray}
\phi^{ABCX}_\star \, \phi^\star_{IJKX} & = & - \frac{3}{8} \, \phi_\star^{AB[IJ} \, \delta^{K]C} \nonumber\\
\null & \null  & - \frac{1}{16} \phi^{ABC} \, \phi_{IJK} \, + \,
\frac{1}{96} \, \delta^{ABC}_{IJK}
\label{fierze}
\end{eqnarray}
which is used to perform the decomposition of the $\mathbf{35}$ representation
of $\mathrm{SO(7)}$ into its irreducible $\mathrm{G_2}$ parts.
\par
The $\mathbf{35}$ representation of $\mathrm{SO(7)}$ is given by an
antisymmetric tensor of rank three. We have:
\begin{equation}
  U^{IJK} \, \sim \, \begin{array}{ccc}
    \null & \null & \yng(1,1,1) \
  \end{array}
\label{35tensor}
\end{equation}
Alternatively, from the $\mathrm{SO(7)}$ point of view, an
antisymmetric tensor can be seen as a traceless, symmetric bispinor.
Indeed, we have that the $8 \times 8$ three-index gamma matrices:
\begin{equation}
  \left(\gamma_{IJK} \right) _{\alpha \beta }
\label{3gamma}
\end{equation}
are symmetric and traceless.
\par
The decomposition of the $\mathbf{35}$ with respect to $\mathrm{G_2} \subset
\mathrm{SO(7)}$ is as follows:
\begin{equation}
  \mathbf{35} \, \stackrel{\mathrm{G_2}}{\longrightarrow} \, \mathbf{27} \, \oplus \, \mathbf{7} \,
  \oplus \, \mathbf{1}
\label{35splitta}
\end{equation}
It is interesting and necessary for our goals to construct the
projectors onto these representations.
\par
The singlet representation of $\mathrm{G_2}$ is given by the
$3$-tensors $U^{IJK}$ proportional to the $\mathrm{G_2}$-invariant $3$--form,
namely:
\begin{equation}
  U^{IJK}_{(1)} \, \propto \, \phi^{IJK}
\label{1repra}
\end{equation}
It is therefore easy to write the projection operator onto the
singlet:
\begin{equation}
  P^{IJK}_{(1)\phantom{IJ}ABC} = \frac{6}{7} \phi_{ABC}\phi^{IJK}
\label{P1pro}
\end{equation}
which has the property :
\begin{equation}
   P^{IJK}_{(1)\phantom{IJ}ABC} \, \phi_{IJK} = \phi_{ABC}
\label{propP1}
\end{equation}
\par
Next we are interested in the projection operator onto the
$\mathbf{7}$--representation. We observe that if we define the two operators:
\begin{equation}
  K_{\mp} = \cases{\delta^{IJK}_{ABC} \, - \, 24 \,
  \phi_\star^{IJKX}\, \phi^{\star}_{ABCX} \cr
  24 \, \phi_\star^{IJKX}\, \phi^{\star}_{ABCX} \cr}
\label{Kpm}
\end{equation}
both of them are projection operators. Indeed:
\begin{equation}
  \left(K_\mp \right)^2 = K_\mp
\label{Kmpsquare}
\end{equation}
Then we observe that defining:
\begin{equation}
  P^{IJK}_{(7)\phantom{IJ}ABC} \, \equiv \, 24 \, \phi_\star^{IJKX}\,
  \phi^{\star}_{ABCX} = K_+
\label{Pon7}
\end{equation}
we have:
\begin{equation}
  P^{IJK}_{(7)\phantom{IJ}ABC} \, P^{XYZ}_{(1)\phantom{IJ}IJK}  \, =
  \, 0
\label{perfetto}
\end{equation}
and we deduce that indeed the operator in eq.(\ref{Pon7}) is the
projection operator onto the $\mathbf{7}$--dimensional
representation.
\par
We consider next the projection operator onto the
$\mathbf{27}$--representation. We set:
\begin{eqnarray}
   P^{IJK}_{(27)\phantom{IJ}ABC} \, = \, 36 \, \phi^{[IJ}_{\star
   \phantom{IJ}[AB} \, \delta^{K]}_{C]} \, + \, 72 \, \phi_\star^{IJKX}\,
  \phi^{\star}_{ABCX} \, + \, \frac{36}{7} \phi_{ABC} \, \phi^{IJK}
\label{P35on27}
\end{eqnarray}
Then we verify that:
\begin{equation}
  P_{(27)}^2 = P_{(27)} \quad ; \quad P_{(27)} \, P_{(1)} = 0 \quad ;
  \quad P_{(27)} \, P_{(1)} = 0
\label{properties1}
\end{equation}
Moreover thanks to the Fierz identity (\ref{fierze}) we have:
\begin{equation}
  P_{(27)} + P_{(7)} + P_{(1)} \, = \, \mathbf{1}
\label{summarulla}
\end{equation}
In this way we have verified that the three projectors perform a
splitting of the $\mathbf{35}$ representation of $\mathrm{SO(7)}$ into three orthogonal
subspaces corresponding to the three irreducible representations of
$\mathrm{G_2}$.
\par
\subsection{Spinorial description of the $\mathbf{27}$ representation}
As we have already recalled the group $\mathrm{G_2}$ is defined as the stability subgroup of an $\mathrm{SO(7)}$
spinor (see eq. (\ref{8in7p1})). The fundamental vector representation of
$\mathrm{SO(7)}$ remains instead irreducible (see eq.(\ref{7in7})).
The same happens for the symmetric tensor representation of $\mathrm{SO(7)}$. A symmetric traceless two-tensor of
$\mathrm{SO(7)}$:
\begin{equation}
  M_{IJ} \, \sim \, \stackrel{\circ}{\yng(2)} \quad ; \quad M_{IJ} =
  M_{JI} \quad ; \quad M_{JJ}=0
\label{27SO7}
\end{equation}
has $27$ independent components and constitutes an irreducible representation.
Under $\mathrm{G_2}$ reduction this representation remains irreducible:
\begin{equation}
  \mathbf{27} \stackrel{G_2}{\longrightarrow} \mathbf{27}
\label{27iirred}
\end{equation}
so that the $\mathbf{27}$ of $\mathrm{G_2}$ has an alternative simpler description in terms of a symmetric traceless tensor. What is the
relation between this and the previous description? It is as usual provided by gamma matrices. Let us see.
\par
The result of our previous discussion is that an irreducible $\mathbf{27}$ representation of $\mathrm{G_2}$ is
provided by an antisymmetric tensor $U^{IJK}_{27}$ satisfying the two $\mathrm{G_2}$ invariant constraints:
\begin{eqnarray}
0 & = & \phi_\star^{ABCT} \, U^{ABC}_{(27)} \nonumber\\
0 & = & \phi^{ABC} \, U^{ABC}_{(27)}
\label{7and1elimina}
\end{eqnarray}
which respectively eliminate the $\mathbf{7}$ and $\mathbf{1}$ irreducible representations.
\par
Let us now consider the spinor index $\alpha = 1,\dots, 8$ of $\mathrm{SO(7)}$ and split it as follows:
\begin{equation}
  \alpha=\cases{A = 1,\dots, 7 \cr 8\cr}
\label{alphasplit}
\end{equation}
which corresponds to the branching rule (\ref{8in7p1}).
Correspondingly, the symmetric, traceless $8 \times 8$ matrices
(\ref{3gamma}) split as follows:
\begin{equation}
  \gamma^{IJK}_{\alpha\beta} = \cases{
  \gamma^{IJK}_{AB} \cr
\gamma^{IJK}_{A8}\cr
\gamma^{IJK}_{88} \cr}
\label{splitta}
\end{equation}
where
\begin{equation}
  \gamma^{IJK}_{AA} = - \gamma^{IJK}_{88}
\label{perdinci}
\end{equation}
follows from $\gamma^{IJK}_{\alpha\beta}$ being traceless.
\par
The decomposition (\ref{splitta}) is the decomposition of the $\mathbf{35}$
representation of $\mathrm{SO(7)}$ into its $\mathbf{27}$, $\mathbf{7}$ and $\mathbf{1}$  components with respect to
$G_2$. Indeed we have that:
\begin{equation}
  \gamma^{IJK}_{88} = 6 \phi^{IJK}
\label{primaagna}
\end{equation}
while:
\begin{equation}
  P^{IJK}_{(7)\phantom{IJ}ABC} \, \gamma^{ABC}_{T8} = \gamma^{IJK}_{T8}
\label{convert7}
\end{equation}
Finally defining the traceless matrices:
\begin{equation}
  W^{IJK}_{XY} \, = \, \gamma^{IJK}_{XY} \, - \, \frac{1}{7} \, \delta_{XY} \gamma^{IJK}_{TT}
\label{ferrati}
\end{equation}
we can verify that:
\begin{equation}
   P^{IJK}_{(27)\phantom{IJ}ABC} \, W^{ABC}_{XY} = W^{IJK}_{XY}
\label{ultraferrati}
\end{equation}
which proves what we just stated.
\subsection{The representation $\mathbf{64}$}
Let us now define the representation $\mathrm{64}$ of $\mathrm{G_2}$.
\par
In the case of $\mathrm{SO(7)}$ we can consider an irreducible traceless tensor with gun symmetry
\begin{equation}
H_{\scriptsize{\begin{array}{cc}
  A & C \\
  B & \null\\
\end{array}}} \, \sim \,  \begin{array}{c}
  \null \\
  \yng(2,1)
\end{array}
\label{olla}
\end{equation}
The tensor $H$ is antisymmetric in $AB$:
\begin{equation}
  H_{\scriptsize{\begin{array}{cc}
  A & C \\
  B & \null\\
\end{array}}}  \, = \, - \,  H_{\scriptsize{\begin{array}{cc}
  B & C \\
  A & \null\\
\end{array}}}
\label{antisym}
\end{equation}
fulfills the cyclic identity:
\begin{equation}
  H_{\scriptsize{\begin{array}{cc}
  A & C \\
  B & \null\\
\end{array}}} \, + \, H_{\scriptsize{\begin{array}{cc}
  C & B \\
  A & \null\\
\end{array}}} \, + \, H_{\scriptsize{\begin{array}{cc}
  B & A \\
  C & \null\\
\end{array}}} \, = \, 0
\label{cyclic}
\end{equation}
and it is traceless:
  \begin{equation}
  H_{\scriptsize{\begin{array}{cc}
  A & B \\
  B & \null\\
\end{array}}} \, = \, 0
\label{stracciato}
\end{equation}
The independent components of such a tensor are $\mathbf{105}$ and
they constitute an irreducible representation of $\mathrm{SO(7)}$. The
branching rule of this representation with respect to $\mathrm{G_2}$ is the
following one:
\begin{equation}
  \mathbf{105} \, \stackrel{\mathrm{G_2}}{\longrightarrow} \, \mathbf{64} \, +
  \, \mathbf{14} \, + \, \mathbf{1}
\label{105split}
\end{equation}
How do we understand it? In the following way. Given the tensor $H$,
by means of the $\mathrm{G_2}$ invariant form we can define the following matrix:
\begin{equation}
  \mathcal{C}_{AB} = \phi_{AIJ} \,  H_{\scriptsize{\begin{array}{cc}
  I & B \\
  J & \null\\
\end{array}}}
\label{vincolmatra}
\end{equation}
Decomposing this matrix into its symmetric and antisymmetric parts:
\begin{eqnarray}
\mathcal{A}_{AB} & \equiv & \frac{1}{2}\left( \mathcal{C}_{AB} \, - \, \mathcal{C}_{BA} \right) \nonumber\\
\mathcal{S}_{AB} & \equiv & \frac{1}{2}\left( \mathcal{C}_{AB} \, + \, \mathcal{C}_{BA} \right)
\label{SandA}
\end{eqnarray}
we can easily verify that:
\begin{eqnarray}
P^{IJ}_{(14)\phantom{IJ}AB} \, \mathcal{A}_{IJ} & = & \mathcal{A}_{AB}  \nonumber\\
{S}_{AA} & = & 0
\label{constonconst}
\end{eqnarray}
This shows that imposing the constraint:
\begin{equation}
  0 \, = \, \phi_{AIJ} \,  H_{\scriptsize{\begin{array}{cc}
  I & B \\
  J & \null\\
\end{array}}}
\label{verlando}
\end{equation}
is just necessary and sufficient to remove the $\mathbf{27}$ and $\mathbf{14}$
irreducible representation of $G_2$ and leave a pure $\mathbf{64}$
representation.
In other words a pure $64$ tensor is a tensor $H_{ABC}$ satisfying
all the constraints (\ref{antisym}), (\ref{cyclic}),
(\ref{stracciato}) and  (\ref{verlando}).
\section{ $7$-dimensional Lie algebras $\mathbb{G}_7$ with $\mathrm{dim} \, \mathrm{Rad} (\mathbb{G}_7) =4$}
\label{Bappendo}
In this appendix we describe the explicit structure of
those $7$--dimensional Lie algebras  whose Levi subalgebra has dimension $3$, or, equivalently the radical has dimension
$4$.
\subsection{Algebras $\mathbb{G}_7$ with Levi subalgebra
$\mathbb{L}(\mathbb{G}_7)=\mathrm{SO(3)}$}
\label{leviso3}
The bosonic representations of $\mathrm{SO(3)}$, namely the
representations $j=n \in \mathbb{Z}_+$ are real, but, due their
dimensionality $2j+1$ they are all odd dimensional. Hence the only
bosonic real representations of $\mathrm{SO(3)}$ with $\mathrm{dim} =4$ are either four
singlets leading to a radical of the following form
\begin{equation}
  \mathrm{Rad}(\mathbb{G}_7) = {\bf 1} \oplus {\bf 1} \oplus {\bf 1} \oplus {\bf 1}
\label{rad7is4sing}
\end{equation}
 or one singlet plus one triplet leading to a radical of the following form:
\begin{equation}
   \mathrm{Rad}(\mathbb{G}_7) = \mathbf{1} \oplus
\mathbf{3}
\label{rad7is1+3}
\end{equation}
\subsubsection{The case of a triplet plus a singlet}
\label{so3w3+1subsecta}
In case (\ref{rad7is1+3})  naming the seven generators as follows:
\begin{equation}
  \left\{J_1, J_2 , J_3 , W_1 , W_2 , W_3 , Z \right\}
\label{JWZorder}
\end{equation}
where $J_x$ are the $\mathrm{SO(3)}$ generators and $W_x$ the triplet
generators, while $Z$ is the singlet, the most general form of the
commutation relations is the following one:
\begin{eqnarray}
\left[ J_x \, , \, J_y \right]  & = & \epsilon_{xyz} \, J_z \nonumber\\
\left[ J_x \, , \, W_y \right]  & = & \epsilon_{xyz} \, W_z \nonumber\\
\left[ J_x \, , \, Z \right]  & = & 0 \nonumber\\
\left[ W_x \, , \, W_y \right]  & = & 0 \nonumber\\
\left[ Z\, , \, W_x \right]  & = & \gamma \, W_x
\label{so3w3+1}
\end{eqnarray}
We name this algebra $\mathbf{so3w3+1}$ and calculating its Ricci
form we obtain:
\begin{equation}
  \mathbf{Ric}_{so3w3+1}= \left( \matrix{ \frac{1}{4} & 0 & 0 & 0 & 0 & 0 & 0 \cr 0 & \frac{1}{4} & 0 & 0 & 0 & 0 & 0 \cr 0 & 0 &
    \frac{1}{4} & 0 & 0 & 0 & 0 \cr 0 & 0 & 0 & \frac{-3\,{\gamma }^2}
   {2} & 0 & 0 & 0 \cr 0 & 0 & 0 & 0 & \frac{-3\,{\gamma }^2}{2} & 0 & 0 \cr 0 & 0 & 0 & 0 & 0 &
    \frac{-3\,{\gamma }^2}{2} & 0 \cr 0 & 0 & 0 & 0 & 0 & 0 & \frac{-3\,{\gamma }^2}{2} \cr  }\right)
\label{ricso3w3+1}
\end{equation}
which has both positive and negative eigenvalues, leading to the
conclusion that this algebra is ruled out as a candidate for weak $\mathrm{G_2}$ holonomy.
\subsubsection{The case of $4$ singlets}
\label{so3w1111}
In this case the algebra $\mathbb{G}_7$ is not the semidirect product of
its Levi subalgebra with its radical but just the \textit{direct}
sum of the same:
\begin{equation}
  \mathbb{G}_7 = \mathbf{SO(3)} \oplus \mathrm{Rad}_(\mathbb{G}_7)
\label{direct}
\end{equation}
Correspondingly the Ricci form is just block diagonal $3+4$. The
contribution of the Levi algebra $\mathbf{SO(3)}$ is just a positive definite matrix,
actually $\ft 14 \times {\bf 1}_{3 \times 3}$ , while, by definition,
$\mathrm{Rad}(\mathbb{G}_7)$ is a completely solvable $4$-dimensional
Lie algebra. Then, for $\mathrm{Rad}(\mathbb{G}_7)$, it is true what in the main text we already proved
for any solvable Lie algebra (see eq.s (\ref{ortoDG+K}), (\ref{squirro} and related text), namely
that the Ricci form has at least one non--positive eigenvalue.
By means of this argument also this case is excluded, leading to an
overall Ricci form non positive definite. It remains to analyze the last case.
\subsubsection{The algebra $\mathbf{so3w4}$}
\label{w4so3}
This Lie algebra is the semidirect product of the rotation algebra
$\mathrm{SO(3)}$ with its only available real $4$-dimensional irreducible
representation, namely the real transcription of the $j=\frac{1}{2}$
spinor representation. The reasoning leading to this algebra is the
following one.  It remains the case of spinor
representations $j=\ft{n}{2}$ with $n \in \mathbb{Z}_+$. These are
all complex representations and they can be transcribed as real
representations in twice their complex dimensions namely in $4j+2$
dimensions. Hence the only possibility is $j=\ft 12$.
\par
To this effect we consider the $4 \times 4$ 't Hooft matrices which
constitute two triplets of either self-dual or antiself dual $\mathrm{SO(3)}$
generators, that commute with each other:
\begin{eqnarray}
\mathbf{J}^{\pm|x}_{ij} & = & - \mathbf{J}^{\pm|x}_{ji} \quad : \quad x=1,2,3 \quad ; \quad i,j= 1,2,3,4 \nonumber\\
\mathbf{J}^{\pm|x}_{ij} & = & \pm \ft 12 \epsilon_{ijkl}\,  \mathbf{J}^{\pm|x}_{kl}
\nonumber\\
\left[ \mathbf{J}^{\pm|x}\, , \, \mathbf{J}^{\pm|y} \right] & = & \epsilon^{xyz} \,
J^{\pm|z} \nonumber\\
\left[ \mathbf{J}^{\pm|x}\, , \, \mathbf{J}^{\mp|y} \right] & = & 0
\label{thofti}
\end{eqnarray}
Then we introduce a basis of generators for the Lie algebra $\mathbf{so3w4}$ named as
follows:
\begin{equation}
  T_I \,  = \, \left\{ J_x \,
  , \, W_i \right \} \, = \, \left\{ J_1,J_2,J_3,W_1,W_2,W_3,W_4 \right\} = \left\{ J_x \,
  , \, W_i \right \}
\label{generiw4so3}
\end{equation}
and the only possible commutation relations are the following ones:
\begin{eqnarray}
\left[ J^x\, , \, J^y \right] & = & \epsilon^{xyz} \,
J^z\nonumber\\
\left[ J^x\, , \, W^i\right] & = & (\alpha \mathbf{J}^{+|x}_{ij} + \beta
\mathbf{J}^{-|x}_{ij}) \, W_j \nonumber\\
\left[ W_i\, , \, W_j \right] & = & 0
\label{so3w4alg}
\end{eqnarray}
where
\begin{equation}
  (\alpha \, , \, \beta) = \cases{(1,0) \, \mbox{or} \, \cr
  (1,1) \, \mbox{or} \, \cr
  (0,1) \, \cr}
\label{perducale}
\end{equation}
Indeed the $4$-dimensional radical has necessarily to be abelian,
since there is no $\mathrm{SO(3)}$ invariant three index tensor in the spinor
representation, or to say it differently the tensor cube of the
$j=\ft 12$ representation does not contain the singlet.
\par
An explicit representation of the 't Hooft matrices is the following
one:
\begin{equation}
  \begin{array}{ccccccc}
    J^{+|1} & = & \left( \matrix{ 0 & \frac{1}{2} & 0 & 0 \cr -  \frac{1}
     {2}    & 0 & 0 & 0 \cr 0 & 0 & 0 & \frac{1}
   {2} \cr 0 & 0 & -  \frac{1}{2}
      & 0 \cr  } \right)  & ; & J^{-|1} & = &\left(  \matrix{ 0 & -
      \frac{1}{2}
      & 0 & 0 \cr \frac{1}
   {2} & 0 & 0 & 0 \cr 0 & 0 & 0 & \frac{1}
   {2} \cr 0 & 0 & - \frac{1}{2}
      & 0 \cr  }\right)  \\
    \null & \null & \null & \null & \null & \null & \null \\
    J^{+|2} & = & \left( \matrix{ 0 & 0 & -  \frac{1}{2}
      & 0 \cr 0 & 0 & 0 & \frac{1}{2} \cr \frac{1}
   {2} & 0 & 0 & 0 \cr 0 & -  \frac{1}{2}
      & 0 & 0 \cr  } \right)  & ; & J^{-|2} & = &\left(  \matrix{ 0 & 0 & \frac{1}{2} & 0 \cr 0 & 0 & 0 & \frac{1}
   {2} \cr - \frac{1}{2}
      & 0 & 0 & 0 \cr 0 & - \frac{1}{2}
      & 0 & 0 \cr  }\right)  \\
    \null & \null & \null & \null & \null & \null & \null \\
    J^{+|3} & = & \left( \matrix{ 0 & 0 & 0 & \frac{1}{2} \cr 0 & 0 & \frac{1}
   {2} & 0 \cr 0 & -  \frac{1}{2}
      & 0 & 0 \cr -  \frac{1}{2}
      & 0 & 0 & 0 \cr  } \right)  & ; & J^{-|31} & = &\left(  \matrix{ 0 & 0 & 0 & -
      \frac{1}{2}
      \cr 0 & 0 & \frac{1}{2} & 0 \cr 0 & -  \frac{1}
     {2}    & 0 & 0 \cr \frac{1}
   {2} & 0 & 0 & 0 \cr  }\right)  \\
    \null & \null & \null & \null & \null & \null & \null \
  \end{array}
\label{thoftoni}
\end{equation}
and it can be used to perform an explicit
calculation of the Ricci form via eq.(\ref{Riccimatrona} ). In the three
cases  provided by eq.s (\ref{so3w4alg}) and
(\ref{perducale}) we obtain three times the same result, namely:
\begin{equation}
  \mathbf{Ric}_{so3w4} = \left(\matrix{ \frac{1}{4} & 0 & 0 & 0 & 0 & 0 & 0 \cr 0 &
    \frac{1}{4} & 0 & 0 & 0 & 0 & 0 \cr 0 & 0 & \frac{1}
   {4} & 0 & 0 & 0 & 0 \cr 0 & 0 & 0 & 0 & 0 & 0 & 0 \cr
   0 & 0 & 0 & 0 & 0 & 0 & 0 \cr 0 & 0 & 0 & 0 & 0 & 0 &
   0 \cr 0 & 0 & 0 & 0 & 0 & 0 & 0 \cr  } \right)
\label{riccionew4so3}
\end{equation}
This shows that for this algebra the Ricci form is degenerate and has
$4$ vanishing eigenvalues. It can be discarded as a candidate for
weak $\mathrm{G_2}$ holonomy.
\subsection{Lie algebras $\mathbb{G}_7$
with Levi subalgebra $\mathbb{L}(\mathbb{G}_7)= \mathbf{so(1,2)}$}
\label{Sl2casi}
These Lie algebras are the semidirect product of the pseudo rotation
algebra $\mathrm{SO(1,2)}$ with a four dimensional solvable Lie
algebra $\mathcal{S}_4$ which must carry a real representation of
$\mathrm{SO(1,2)}$. Here the situation is different from the case of
the compact Levi algebra $\mathrm{SO(3)}$ since the two dimensional
representation of $\mathrm{SO(1,2)}$ can be chosen real and identified with
the defining representation of $\mathrm{SL(2,\mathbb{R})}$. Hence we
have the following subcases, depending on the representation
assignments of the $\mathcal{S}_4$ radical:
\begin{description}
\item [a)] $\mathcal{S}_4 = \mathbf{4} $
 \item [b)] $\mathcal{S}_4 = \mathbf{2} \oplus  \mathbf{2}$
  \item [c)] $\mathcal{S}_4 = \mathbf{2} \oplus \mathbf{1} \oplus \mathbf{1} $
  \item [d)] $\mathcal{S}_4 = \mathbf{1} \oplus \mathbf{1} \oplus \mathbf{1} \oplus \mathbf{1} $
\end{description}
where $\mathbf{1}$ denotes the singlet representation, $\mathbf{2}$
denotes the $j=\ft 12 $ representation and $\mathbf{4}$ denotes the
irreducible $j=\ft 32 $ representation. Let us discuss these four
cases separately.
\subsubsection{a) The irreducible case a}
Here the structure of the Lie algebra is completely fixed. Let
\begin{equation}
\begin{array}{rcl}
  \Lambda_1 & = & \left(\matrix{ \frac{3}{2} & 0 & 0 & 0 \cr 0 & \frac{1}
   {2} & 0 & 0 \cr 0 & 0 & - \frac{1}{2}
      & 0 \cr 0 & 0 & 0 & - \frac{3}{2}   \cr  } \right)  \\
      \null & \null & \null \\
  \Lambda_2 & = & \left(\matrix{ 0 & \frac{3}{2} & 0 & 0 \cr \frac{1}
   {2} & 0 & 1 & 0 \cr 0 & 1 & 0 & \frac{1}{2} \cr 0 & 0 & \frac{3}
   {2} & 0 \cr  } \right)  \\
    \null & \null & \null \\
  \Lambda_3 & = & \left(\matrix{ 0 & \frac{3}{2} & 0 & 0 \cr - \frac{1}{2}
      & 0 & 1 & 0 \cr 0 & -1 & 0 & \frac{1}{2} \cr 0 & 0 & -
     \frac{3}{2} \  & 0 \cr }  \right)
\end{array}
\label{tremezgener}
\end{equation}
be the generators of $\mathrm{SO(1,2)}$ in the irreducible $j=\ft 32$
representation which corresponds to the $3$-times symmetric tensor product
of the fundamental $j=1$ representation:
\begin{equation}
  \begin{array}{ccc}
    \lambda_1 & = & \left( \matrix{ \frac{1}{2} & 0 \cr 0 & - \frac{1}{2}   \cr  }\right)  \\
     \null & \null & \null \\
    \lambda_2 & = & \left( \matrix{ 0 & \frac{1}{2} \cr \frac{1}{2} & 0 \cr  } \right) \\
     \null & \null & \null \\
    \lambda_3 & = &\left(  \matrix{ 0 & \frac{1}{2} \cr - \frac{1}{2}   & 0 \cr  } \right) \
  \end{array}
\label{unmezzo}
\end{equation}
Naming $f^x_{\phantom{x}yz}$ the structure constants of the $\mathrm{SO(1,2)}$
simple Lie algebra:
\begin{equation}
\left[   J_y \, , \, J_z \right] \, = \, f^x_{\phantom{x}yz} \,
J_z \quad ; \quad \left[   \Lambda_y \, , \, \Lambda_z \right] \, = \, f^x_{\phantom{x}yz} \,
\Lambda_z \quad ; \quad \left[   \lambda_y \, , \, \lambda_z \right] \, = \, f^x_{\phantom{x}yz} \,
\lambda_z
\label{so12commu}
\end{equation}
the only Lie possible Lie algebra corresponding to this case is given
by:
\begin{eqnarray}
\left[ J_x\, , \, J_y \right] & = & f^x_{\phantom{x}yz} \,
J^z\nonumber\\
\left[ J_x\, , \, W_i\right] & = & \left( \Lambda_z \right)_i^{\phantom{i}j}  \, W_j \nonumber\\
\left[ W_i\, , \, W_j \right] & = & 0
\label{so12w4alg}
\end{eqnarray}
where $J_x$ ($x=1,2,3$) are the generators of $\mathrm{SO(1,2)}$ and
$W_i$ ($i=1,2,3,4$) are the generators spanning the $j=\ft 32$
representation. We name the above algebra $\mathbf{so12w4}$. If we order the
seven generators in the following way:
\begin{equation}
  \left\{ J_1 , J_2, J_3 , W_1 , W_2 , W _3 , W_4 \right \}
\label{ordinew4so12}
\end{equation}
we can evaluating the Ricci
form by means of the formula (\ref{Riccimatrona}) and  we find:
\begin{equation}
\mathbf{Ric}_{so12w4} =   \left( \matrix{ - \frac{13}{4}
      & 0 & 0 & 0 & 0 & 0 & 0 \cr 0 & - \frac{15}{4}
      & 0 & 0 & 0 & 0 & 0 \cr 0 & 0 & - \frac{1}{4}
      & 0 & 0 & 0 & 0 \cr 0 & 0 & 0 & 1 & 0 & 0 & 0 \cr 0 & 0 & 0 &
   0 & -1 & 0 & 0 \cr 0 & 0 & 0 & 0 & 0 &
    -1 & 0 \cr 0 & 0 & 0 & 0 & 0 & 0 & 1 \cr  }\right)
\label{ricciso12w4}
\end{equation}
As we see, also in this case the Ricci form has both positive and
negative eigenvalues, so also $\mathbf{w4so12}$ is ruled out as a candidate
for weak $\mathrm{G_2}$ holonomy.
\subsubsection{b) The case of two doublets}
In this case the Lie algebra is also completely fixed by the choice
of the representations. The seven generators are arranged into one
triplet $J_x$ ($x=1,2,3$), corresponding to the adjoint representation and two
doublets $W_\alpha$ and $U_\alpha$ ($\alpha = 1,2$). The commutation
relations are:
\begin{eqnarray}
\left[ J_x\, , \, J_y \right] & = & f^x_{\phantom{x}yz} \,
J^z\nonumber\\
\left[ J_x\, , \, W_\alpha\right] & = & \left( \lambda_z \right)_\alpha^{\phantom{i}\beta}  \, W_\beta \nonumber\\
\left[ J_x\, , \, U_\alpha\right] & = & \left( \lambda_z \right)_\alpha^{\phantom{i}\beta}  \, U_\beta \nonumber\\
\left[ W_\alpha\, , \, W_\beta \right] & = & 0 \nonumber\\
\left[ U_\alpha\, , \, U_\beta \right] & = & 0 \nonumber\\
\left[ W_\alpha\, , \, U_\beta \right] & = & 0 \nonumber\\
\label{so12w2u2alg}
\end{eqnarray}
where the matrices $\lambda_x$ were defined in eq. (\ref{unmezzo})
and the $\mathrm{SO(1,2)}$ structure constants in (\ref{so12commu}).
We name the above algebra $\mathbf{so12w2u2}$. Arranging the
generators in the following order:
\begin{equation}
  \left\{ J_1 , J_2, J_3 , W_1 , W_2 , U _1 , U_2 \right \}
\label{ordinew2u2so12}
\end{equation}
The calculation of the Ricci form yields:
\begin{equation}
  \mathbf{Ric}_{so12w2u2} =   \left( \matrix{ - \frac{5}{4}
      & 0 & 0 & 0 & 0 & 0 & 0 \cr 0 & - \frac{5}{4}
      & 0 & 0 & 0 & 0 & 0 \cr 0 & 0 & \frac{1}
   {4} & 0 & 0 & 0 & 0 \cr 0 & 0 & 0 & 0 & 0 & 0 & 0 \cr 0 & 0 & 0 &
   0 & 0 & 0 & 0 \cr 0 & 0 & 0 & 0 & 0 & 0 & 0 \cr 0 & 0 & 0 & 0 & 0 &
   0 & 0 \cr  }\right)
\label{ricciso12w2u2}
\end{equation}
We have positive, negative and null eigenvalues. Hence also this
algebra is ruled out.
\subsubsection{c) The case of one doublet and two singlets.}
With these representation assignments there is more than one algebra
which is possible. So we have to study a few subcases. Let us name
$J_x$ the triplet generators and $W_\alpha$ the doublet ones. The
remaining two, we can call $Z$ and $U$. The following commutation
relations are fixed by the representation assignments:
\begin{eqnarray}
\left[ J_x\, , \, J_y \right] & = & f^x_{\phantom{x}yz} \,
J^z\nonumber\\
\left[ J_x\, , \, W_\alpha\right] & = & \left( \lambda_z \right)_\alpha^{\phantom{i}\beta}  \, W_\beta \nonumber\\
\left[ J_x\, , \, U\right ] & = & 0 \nonumber\\
\left[ J_x\, , \, Z \right ] & = & 0 \nonumber\\
\label{so12w2uzalg1}
\end{eqnarray}
For the remaining commutators we have a bifurcation. We can decide
that the doublet generators are not abelian, rather, together with
one of the singlet, let us say $Z$, they form a nilpotent algebra.
This is possible because the antisymmetric square of the $j=\ft 12$
representation contains the singlet. Hence we can write:
\begin{equation}
  \left[ W_\alpha\, , \, W_\beta \right] = \epsilon^{\alpha\beta} \,
  Z
\label{so12w2uzalg2}
\end{equation}
If we choose this path then the most general commutation relations
with the last generator $U$ are fixed by Jacobi identities and are the
following one:
\begin{eqnarray}
\left[ U\, , \, W_\alpha \right ] & = & q \, W_\alpha \nonumber\\
\left[ U\, , \, Z \right ] & = & 2 \, q \, Z \nonumber\\
\label{so12w2uzalg3}
\end{eqnarray}
The algebra defined by the commutation relations
(\ref{so12w2uzalg1},\ref{so12w2uzalg2},\ref{so12w2uzalg3}) we name
$\mathbf{so12w2uz}$ and we can calculate its Ricci form by arranging its
generators in the following order:
\begin{equation}
  \left\{ J_1 , J_2, J_3 , W_1 , W_2 , Z , U \right \}
\label{ordineso12w2uz}
\end{equation}
The result is:
\begin{equation}
  \mathbf{Ric}_{so12w2uz} \, = \, \left(\matrix{ -1 & 0 & 0 & 0 & 0 & 0 & 0 \cr 0 &
    -1 & 0 & 0 & 0 & 0 & 0 \cr 0 & 0 & \frac{1}
   {4} & 0 & 0 & 0 & 0 \cr 0 & 0 & 0 & - \frac{1}{4}   -
   2\,q^2 & 0 & 0 & 0 \cr 0 & 0 & 0 & 0 & - \frac{1}{4}
       - 2\,q^2 & 0 & 0 \cr 0 & 0 & 0 & 0 & 0 & \frac{1}{4} -
   4\,q^2 & 0 \cr 0 & 0 & 0 & 0 & 0 & 0 & -3\,q^2 \cr  } \right)
\label{ricciso12w2uz}
\end{equation}
By explicit inspection we see that irrespectively of the value of
$q$, the signature of the Ricci form is indefinite including both
positive and negative eigenvalues. Hence also this case has to be
discarded.
\par
The other possibility for a Lie algebra with the chosen
representation assignments is realized by keeping
eq.(\ref{so12w2uzalg1}) and replacing eq. (\ref{so12w2uzalg2}) with:
\begin{equation}
  \left[ W_\alpha\, , \, W_\beta \right] = 0
\label{so12w2uzalg2bis}
\end{equation}
At the same time we can replace eq.(\ref{so12w2uzalg3}) with
\begin{eqnarray}
\left[ U\, , \, W_\alpha \right ] & = & q_2 \, W_\alpha \nonumber\\
\left[ Z\, , \, W_\alpha \right ] & = & q_1 \, W_\alpha \nonumber\\
\left[ U\, , \, Z \right ] & = & 0
\label{so12w2uzalg3bis}
\end{eqnarray}
The algebra defined by
eq.s(\ref{so12w2uzalg1},\ref{so12w2uzalg2bis},\ref{so12w2uzalg3bis}),
we name $\mathbf{so12w2q1q2}$. We can calculate its Ricci form by arranging its
generators in the same order (\ref{ordineso12w2uz}) as above. The result is given below:
\begin{eqnarray}
\mathbf{Ric}_{so12w2q1q2} & = & \left(\matrix{ -1 & 0 & 0 & 0 & 0 & 0 & 0 \cr 0 &
    -1 & 0 & 0 & 0 & 0 & 0 \cr 0 & 0 & \frac{1}
   {4} & 0 & 0 & 0 & 0 \cr 0 & 0 & 0 & -{{q_1}}^2 -
   {{q_2}}^2 & 0 & 0 & 0 \cr 0 & 0 & 0 & 0 & -{{q_1}}^2 -
   {{q_2}}^2 & 0 & 0 \cr 0 & 0 & 0 & 0 & 0 & -{{q_1}}^2 & -\left(
     {q_1}\,{q_2} \right)  \cr 0 & 0 & 0 & 0 & 0 & -\left( {q_1}\,
     {q_2} \right)  & -{{q_2}}^2 \cr  } \right)  \nonumber\\
\label{so12w2q1q2}
\end{eqnarray}
As we see also this Ricci form has indefinite signature. It has both
positive and negative eigenvalues and also a null eigenvalue. Hence
also this case has to be discarded.
\subsubsection{d) The case of four singlets}
This case is easily disposed of. If four of the seven generators are
singlets, then it means that our algebra is actually the direct
product of the Levi subalgebra $\mathrm{SO(1,2)}$ which by itself has
already an indefinite Ricci form with a completely solvable
subalgebra for which the Ricci form has at least one negative
eigenvalue as we have proved in the main text. Indeed for direct
product algebras the Ricci form is obviously block diagonal.
So all such cases are automatically excluded.

\end{document}